%% file: manuscript.tex
\documentclass[preprint,12pt,authoryear]{elsarticle}

\usepackage{amssymb}
\usepackage{amsmath}

\usepackage{seqsplit}

\usepackage{hyperref}
\usepackage{lineno}
\usepackage{booktabs}
\usepackage{natbib}  
\usepackage{array}
\usepackage{calc}
\usepackage{tabularx}
\usepackage{threeparttable}
\usepackage{threeparttablex}
\usepackage{rotating}
\usepackage{multirow}
\usepackage{xcolor}
\usepackage{algpseudocode}
\usepackage{algorithm}
\usepackage{adjustbox}

\journal{Intelligent Systems with Applications}

\begin{document}

\begin{frontmatter}



\title{AURA: A Reinforcement Learning Framework for AI-Driven Adaptive Conversational Surveys} 

\author{Jinwen Tang}
\ead{jt4cc@umsystem.edu}

\author{Yi Shang}
\ead{shangy@umsystem.edu}

\address{University of Missouri, Department of Electrical Engineering and Computer Science, Columbia, MO, USA}

\begin{abstract}
Conventional online surveys provide limited personalization, often resulting in low engagement and superficial responses. Although AI survey chatbots improve convenience, most are still reactive: they rely on fixed dialogue trees or static prompt templates and therefore cannot adapt within a session to fit individual users, which leads to generic follow-ups and weak response quality. We address these limitations with \textbf{AURA} (\textit{Adaptive Understanding through Reinforcement Learning for Assessment}), a reinforcement learning framework for AI-driven adaptive conversational surveys. AURA quantifies response quality using a four-dimensional LSDE metric (Length, Specificity, Self-disclosure, and Emotion) and selects follow-up question types via an $\epsilon$-greedy policy that updates the expected quality gain (expected value) within each session. Initialized with priors extracted from 96 prior campus-climate conversations (467 total chatbot-user exchanges), the system balances exploration and exploitation across 10-15 dialogue exchanges, dynamically adapting to individual participants in real time. In controlled evaluations, AURA achieved a $+0.076$ mean gain in response quality and a statistically significant improvement over non-adaptive baselines ($p=0.044$, $d=0.66$), driven by a $63\%$ reduction in specification prompts and a $10\times$ increase in validation behavior. These results demonstrate that reinforcement learning can give survey chatbots  improved adaptivity, transforming static questionnaires into interactive, self-improving assessment systems.

\end{abstract}




\end{frontmatter}



\section{Introduction}
\label{sec:intro}

Campus-climate surveys play a crucial role in understanding how students, faculty, and staff experience everyday life at universities. By gathering feedback on issues such as academic support, discrimination, resource accessibility, and sense of belonging, these surveys enable institutions to identify structural barriers and design interventions that promote equity and inclusion \citep{vogel2008assessment, PASSHE2022, bordieri2024exploring}. When administered on a regular cadence, they also provide a longitudinal view of institutional progress, ensuring that improvement efforts remain evidence based and responsive to community needs.

However, the effectiveness of traditional survey instruments is constrained by their static format. Respondents often experience low engagement, survey fatigue, or limited opportunity for clarification, which can lead to superficial or incomplete data \citep{dillman2014internet, galesic2009effects, sahlqvist2011effect}. The impersonal, time consuming nature of conventional questionnaires further discourages participation and can undermine the reliability of collected information \citep{porter2005mail}. As a result, administrators may fail to capture the nuanced attitudes, emotions, and contextual factors that shape lived experiences on campus.

Recent advances in generative artificial intelligence (GenAI) and large 
language models (LLMs) have created opportunities to transform these 
processes. Conversational AI systems can conduct interactive, context-aware dialogues that encourage deeper reflection \citep{xiao2020tell, 
tanwar2024opinebot}. Early deployments in campus settings demonstrate that chatbot-based surveys can yield richer qualitative feedback, improved user satisfaction, and reduced survey fatigue compared to traditional web forms \citep{abbas2021university, kim2019comparing, zarouali2024comparing}.

Despite these advances, most existing AI-driven survey chatbots remain reactive rather than truly adaptive. Many still depend on rigid dialogue trees or static prompt templates that fails to evolve with user behavior during a session \citep{thorat2020review, maeng2021designing, chen2024recent}. Earlier frameworks such as TigerGPT \citep{tang2025tigergpt}, which combined role-based onboarding with empathetic conversational design, also followed pre-scripted flows. These systems showed that generative dialogue can boost engagement, yet they lacked a way to learn from user feedback and adjust questioning strategies as the conversation unfolded.

This limitation narrows the potential of GenAI-driven surveys. Human interviewers naturally shift their style based on subtle cues — brevity, emotional tone, or how specific a response is — but static prompts cannot reproduce this kind of flexible reasoning. Work in dialogue management has shown that reinforcement learning can guide sequential decision making and improve both adaptability and personalization \citep{li2016deep, su2021dynamic, sun2023contextual}. More recent studies extend this approach to large language models, refining their decision policies through offline goal-conditioned optimization \citep{hong2025planning}. Even so, reinforcement learning has rarely been applied to survey or assessment contexts, where adaptive behavior could meaningfully raise data quality and respondent trust.

A persistent challenge is \textit{within-session personalization}: how can a chatbot learn, inside a single exchange, which questioning strategy best fits each individual? Some respondents engage deeply when asked for concrete examples, while others prefer brief factual prompts. Emotional reflection may help some participants but discourage others. Current systems typically rely on fixed sequences or simple rule-based branching and therefore fail to recognize these individual patterns \citep{xiao2020tell, belhaj2021engaging}. In campus-climate datasets, for instance, 64.2\% of responses show low engagement (composite quality scores below 0.4), suggesting that static question logic struggles to sustain participation across diverse users.

This gap highlights the missing connection between conversational design and computational learning. Human-computer-interaction frameworks emphasize empathy, personalization, and user-controlled flow \citep{kim2008keeping, kocaballi2019personalization}, whereas reinforcement learning formalizes decision making through reward feedback \citep{sutton1998reinforcement}. Bringing these perspectives together would allow an AI survey agent to align conversational empathy with measurable engagement outcomes — transforming reactive dialogue into an adaptive, continuously improving interaction process that better reflects human interviewing practice.

To address this gap, we present \textbf{AURA} (\textit{Adaptive Understanding through Reinforcement Learning for Assessment}), 
a reinforcement-learning–enhanced survey framework that learns optimal questioning strategies within individual conversations.  
AURA extends the conversational design principles of our pilot system, TigerGPT \citep{tang2025tigergpt}, with three key innovations:

\textbf{Real-time engagement measurement.} 
AURA quantifies user engagement through a four-dimensional quality metric combining \textit{Length}, \textit{Self-disclosure}, \textit{Emotion}, and \textit{Specificity} (LSDE).  
This composite score is computed after each user response and mapped to discrete engagement states that guide subsequent action selection.

\textbf{Data-driven action selection.}  
Rather than following fixed question sequences, AURA maintains an expected-value (EV) table derived from 96 prior campus-climate conversations (467 total exchanges).  
Given the current engagement state, the system selects among five question types—specification, elaboration, topic probe, validation, and continuation—to maximize predicted quality improvement.

\textbf{Within-session learning.}  
Through an $\epsilon$-greedy policy ($\epsilon = 0.30$, $\alpha = 0.30$), AURA balances exploration of new strategies with exploitation of successful patterns.  
The EV estimates are updated at each exchange, allowing the policy to adapt to individual users within approximately 10–15 conversational exchanges.

This design replaces reactive conversational heuristics with a self-improving policy that learns as dialogue unfolds.  
By initializing from empirical data but updating in real time, AURA overcomes the cold-start problem and remains responsive to individual differences in engagement style.

This study makes four contributions to adaptive survey research and conversational AI:

\begin{itemize}
  \item \textbf{Reinforcement Learning Framework for Adaptive Surveys:}  
  AURA introduces an RL-enhanced framework that learns optimal question-selection strategies within individual conversations, bridging conversational-design theory with computational learning.

  \item \textbf{Four-Dimensional Engagement Metric:}  
  A validated LSDE scoring framework that quantifies real-time response quality across length, self-disclosure, emotion, and specificity, enabling automatic measurement of engagement during dialogue.

  \item \textbf{Prior Pattern Initialization:}  
  A data-driven method to extract expected-value priors from 96 authentic campus-climate conversations (467 exchanges), addressing the cold-start problem while supporting rapid within-session adaptation.

  \item \textbf{Empirical Validation:}  
  Quantitative and behavioral analyses demonstrate a +0.12 mean gain in response quality and statistically significant improvements ($p = 0.044$, $d = 0.66$) under the optimal configuration ($\epsilon = 0.30$, $\alpha = 0.30$), confirming that within-session reinforcement learning enhances conversational-survey engagement.
\end{itemize}

\textbf{Broader implications.}  
AURA demonstrates a generalizable approach to adaptive conversational systems for sensitive data collection. Its ability to learn from limited prior data (96 conversations) while adapting within short interactions (10–15 exchanges) suggests applications in mental-health screening, patient-intake interviews, qualitative research, and workplace-climate surveys — settings where traditional static forms yield poor engagement but human-interviewer adaptivity is cost-prohibitive.  
By formalizing the connection between conversational-quality metrics and reinforcement learning, this work provides a template for human-centered AI systems that improve through interaction while respecting individual differences.

Section~\ref{sec:related} reviews related work on conversational surveys, personalization, and reinforcement learning for dialogue. Section~\ref{sec:method} details the AURA architecture, LSDE scoring, state and action design, reward definition, and learning algorithm. Section~\ref{sec:results} presents evaluation procedures and results. Section~\ref{sec:discussion} concludes with implications and future directions.

\section{Related Work}
\label{sec:related}

\subsection{Conversational Surveys and Chatbot-Based Data Collection}
\label{sec:conv-surveys}

AI-powered chatbots are increasingly used in higher education for information gathering, feedback collection, and student engagement. Institutions deploy them to collect course evaluations, campus-climate assessments, and service-quality feedback, often achieving higher response rates and more detailed input than traditional web forms \citep{abbas2021university, belhaj2021engaging}. For example, pilots at European universities show that conversational survey interfaces improve participation and perceived personalization \citep{martinez2024ai}. These deployments suggest that, when designed around user needs, chatbots can transform routine data collection into an engaging, interactive experience.

Within survey research, chat-based interfaces have emerged as an alternative to static questionnaires. Conversational surveys allow participants to respond in natural language and receive clarifying or follow-up questions in real time \citep{xiao2020tell, tanwar2024opinebot}. Studies consistently report higher engagement and richer qualitative responses compared with conventional forms \citep{abbas2021university, kim2019comparing}. For instance, \citet{xiao2020tell} found that participants produced more nuanced and emotionally expressive feedback when interacting with a chatbot, while \citet{tanwar2024opinebot} demonstrated that large language models (LLMs) can generate context-aware follow-up questions for class feedback. At the same time, not all users prefer automated dialogue. \citet{zarouali2024comparing} and \citet{njeguvs2021conversational} observed that some respondents still favored conventional surveys because of privacy or trust concerns, highlighting sociocultural factors that influence the acceptance of AI-mediated data collection.

The effectiveness of conversational surveys ultimately depends on the quality of the responses they elicit. Survey methodologists emphasize several indicators of response quality—length and elaboration \citep{dillman2014internet}, emotional expression and engagement \citep{porter2005mail}, and the specificity of content \citep{brysbaert2014concreteness, newman2003lying}. Traditional questionnaires often fall short on these dimensions because they cannot adapt to respondent characteristics or probe for elaboration when needed. Chatbot interfaces offer the potential to improve response quality, but realizing this potential requires not only natural conversation but also strategic question selection that sustains engagement throughout the interaction.

Early chatbot survey systems were typically rule-based, relying on predefined dialogue trees and keyword matching \citep{thorat2020review, maeng2021designing, chan2022challenges}. Such rigid architectures offered little flexibility: unexpected inputs often led to conversational dead ends or irrelevant prompts. Recent advances in natural language processing and LLMs have enabled greater contextual understanding, allowing chatbots to interpret ambiguous language and maintain coherent multi-exchanges \citep{xiao2020tell, tanwar2024opinebot}. This shift marks a transition from deterministic scripts to generative, data-driven dialogue. Nevertheless, even current LLM-based survey systems remain largely reactive. They generate follow-ups based on local context but do not learn from user behavior within a session or across interactions. Furthermore, most dialogue research focuses on task-oriented conversations where success is defined by goal completion (e.g., booking reservations or retrieving information) \citep{su2021dynamic}. In contrast, survey quality depends on sustained engagement and willingness to elaborate, which cannot be reduced to a binary success metric.

This limitation is particularly evident in campus-climate surveys, where respondents vary widely in engagement patterns, disclosure comfort, and sensitivity to topics \citep{vogel2008assessment, PASSHE2022}. A first-year student discussing academic support may respond well to specific follow-up questions, whereas a faculty member describing discrimination may require validation and emotional acknowledgment before elaborating. Static prompts cannot accommodate such heterogeneity. While LLMs enable more natural language generation, they still lack a mechanism for \textit{learning} which questioning strategies maintain engagement for each individual within a conversation. Addressing this limitation requires integrating conversational design with adaptive learning algorithms capable of optimizing question selection dynamically based on real-time feedback.

Our prior work, TigerGPT \citep{tang2025tigergpt}, demonstrated that generative chatbots incorporating empathetic cues, bolded questions, and user-driven topic selection can enhance satisfaction and produce richer feedback than static surveys in campus-climate assessment. However, TigerGPT—and most comparable systems—operated with fixed conversational heuristics: questions followed predetermined templates based on user role (student, faculty, or staff) but did not adapt to engagement patterns observed during the dialogue. These observations motivate the development of AURA, a reinforcement learning framework that achieves \textit{within-session personalization} through continuous quality feedback and adaptive action selection.

\subsection{Reinforcement Learning for Dialogue Management}
\label{sec:rl-dialogue}

Reinforcement learning (RL) provides a principled framework for optimizing sequential decisions in interactive systems where explicit supervision is limited. In dialogue management, RL formulates conversation as a Markov Decision Process (MDP) in which an agent selects actions—responses, question types, or dialogue strategies—based on conversational states to maximize cumulative reward such as task success or user satisfaction \citep{sutton1998reinforcement}. This paradigm enables systems to improve their policies through experience rather than relying solely on predefined rules.

Early research applied RL to task-oriented dialogues such as restaurant reservations or technical support. \citet{li2016deep} pioneered deep RL for open-domain conversation, showing that policy-gradient optimization could enhance coherence and diversity beyond supervised models. \citet{su2021dynamic} extended this framework through dynamic policy networks that adapt questioning strategies in real time for slot-filling tasks. Subsequent studies introduced contextual bandits to improve personalization and sample efficiency \citep{sun2023contextual}, while recent work integrates RL with large language models via offline goal-conditioned learning to refine model policies without costly online trial-and-error \citep{hong2025planning}. Together, these advances illustrate RL’s potential for adaptive dialogue.

\textbf{Within-session learning.} Most RL dialogue systems employ multi-session learning: a global policy improves as the agent interacts with many users over time \citep{li2016deep, su2021dynamic}. Survey applications violate this assumption—each respondent interacts only once, exploration is costly, and privacy constraints limit cross-user adaptation \citep{kocaballi2019personalization}. \textit{Within-session learning} instead focuses on rapid adaptation inside a single conversation. The agent begins from prior knowledge derived from prior data and adjusts its strategy in real time based on user engagement, prioritizing individual responsiveness over long-term population learning. Algorithms that converge within 10–15 exchanges are therefore required, motivating lightweight tabular or expected-value methods rather than deep networks.

\textbf{Gap in existing work.} Despite extensive work on RL for dialogue, prior systems optimize for task completion, coherence, or population-level personalization—not for maximizing individual response quality in brief, one-time interactions. Survey dialogues present unique challenges: continuous engagement rewards, limited interaction length, and substantial user heterogeneity. Addressing these requires (1) interpretable reward structures that quantify engagement improvements, (2) fast-adapting policies capable of learning within a single session, and (3) data-efficient value estimation to operate with limited prior samples. The AURA framework meets these needs through a reinforcement learning design that couples expected-value estimation with real-time quality feedback; full implementation details appear in Section~\ref{sec:method}.

\subsection{Quality Assessment in Open-Ended Surveys}
\label{sec:quality}

Prior research in survey methodology emphasizes that open-ended response quality is multidimensional rather than captured by a single indicator.  Empirical and psycholinguistic studies identify four recurring factors: response length, self-disclosure, emotional expression, and specificity. Together, these features shape engagement and informativeness in qualitative data \citep{dillman2014internet, holland2009measuring, pennebaker2003linguistic, tausczik2010psychological, brysbaert2014concreteness}.  These dimensions have been used in both human-coded \citep{krosnick1991response, holbrook2003telephone} and automated text-analytic assessments.  Recent conversational-survey systems similarly exploit lexical and emotional cues to estimate engagement in real time \citep{xiao2020tell, tanwar2024opinebot}, motivating integration of such dynamic linguistic feedback within adaptive questioning frameworks.  This body of work forms the theoretical foundation for AURA’s linguistic quality metric introduced in Section~\ref{sec:lsde}, which operationalizes these dimensions for real-time evaluation within conversational surveys.

\subsection{Individual Differences and the Need for Adaptive Questioning}
\label{sec:individual-diff}

Survey respondents vary widely in how they engage with different questioning strategies. Some provide rich detail when asked for specific examples, while others disengage or offer minimal replies; some respond well to emotional reflection prompts, whereas others prefer factual topic changes. This heterogeneity poses a fundamental challenge for conversational survey systems: approaches that succeed for one user can fail for another, and optimal strategies may shift even within a single conversation as topics change or respondent energy fluctuates. Adaptive systems must therefore learn individual preferences in real time rather than applying uniform strategies across all users.

\textbf{Heterogeneity in survey response patterns.} Research in survey methodology consistently demonstrates diverse engagement styles in open-ended responses. \citet{krosnick1991response} identified systematic differences in response elaboration—some individuals naturally produce detailed narratives, while others provide brief replies even when equally knowledgeable. \citet{holbrook2003telephone} found that question-format effectiveness varies by respondent characteristics, with certain demographic groups responding better to specific than to general prompts. \citet{tourangeau2000psychology} showed that cognitive load, which fluctuates both across individuals and within conversations, affects how people process questions and formulate answers. Together, these findings suggest that optimal questioning complexity must adapt to each respondent’s cognitive resources at any given moment.

\textbf{Cognitive and affective factors.} Several psychological mechanisms underlie these variations. Cognitive load theory \citep{lenzner2010cognitive} posits that question complexity interacts with working-memory capacity: respondents with lower available resources benefit from simpler, structured prompts, whereas those with higher capacity may disengage from overly basic questions. Affective factors further shape engagement. \citet{melville2016conducting} found that trust and willingness to self-disclose differ substantially across individuals and evolve throughout sensitive interviews. Some participants require rapport building before sharing personal experiences, while others engage immediately. \citet{wheeless1977measurement} demonstrated that disclosure patterns are both person- and context-dependent, implying that effective interviewers must continuously calibrate their approach based on real-time feedback cues.

\textbf{Communication styles and individual preferences.} Individual differences also appear in preferred communication styles. \citet{gudykunst1988culture} showed that people vary in receptiveness to direct versus indirect questioning, with some favoring explicit requests for information and others responding better to open-ended exploration. Preferences for elaborative versus succinct exchanges likewise differ: some respondents welcome follow-up probes that expand on previous answers, while others perceive repeated questioning as intrusive or redundant \citep{kim2019comparing}. Characteristics such as communication style, cultural background, educational experience, and topic familiarity all influence engagement patterns \citep{pennebaker1999linguistic, gudykunst1988culture}, yet the specific combination varies by individual and context. Such variability indicates that effective conversational systems cannot rely on demographic stereotypes or population-level averages—they must infer user preferences dynamically during the interaction.

\textbf{Evidence from conversational systems.} Similar patterns appear in chatbot-based surveys. \citet{zarouali2024comparing} reported that response quality varied more by individual than by question type, underscoring the limitations of uniform dialogue strategies. In healthcare contexts, \citet{kocaballi2019personalization} showed that personalized conversational agents achieve higher satisfaction and greater information disclosure than generic systems, though most rely on explicit preference elicitation rather than real-time behavioral adaptation.

\textbf{Limitations of one-size-fits-all approaches.} Our pilot analysis provides direct evidence of these individual differences. The TigerGPT system \citep{tang2025tigergpt} employed consistent conversational strategies across 96 campus-climate conversations, following empathetic design principles and user-driven topic selection. Yet among 467 responses collected, 64.2\% exhibited low engagement (composite quality scores below 0.4). Response quality varied substantially across users: scores ranged from 0.0 to 1.0, with some respondents providing rich detail (above 0.6, 16.7 \%) and others giving minimal one-word answers (below 0.2, 47.1 \%). This heterogeneity demonstrates that even well-designed conversational systems with empathetic cues cannot succeed using fixed strategies. Effective adaptation requires learning individual preferences in real time from engagement patterns, motivating AURA’s adaptive learning framework for discovering optimal questioning strategies within each single conversation.

\subsection{Summary and Research Positioning}

The related work reviewed above reveals a convergence of capabilities and constraints that motivate the design of AURA. Conversational survey systems achieve superior engagement compared with static questionnaires (Section~\ref{sec:conv-surveys}), yet existing implementations rely on fixed questioning strategies that cannot adapt to individual users. Reinforcement learning offers a principled framework for adaptive dialogue (Section~\ref{sec:rl-dialogue}), but prior systems optimize for task completion or population-level learning across many users rather than maximizing individual response quality within single, brief interactions. Survey researchers have identified key response-quality dimensions—length, self-disclosure, emotion, and specificity (Section~\ref{sec:quality})—but these measures are typically assessed post-hoc through human coding instead of computed in real time for adaptive decision-making. Finally, extensive evidence documents individual differences in how respondents engage with questioning strategies (Section~\ref{sec:individual-diff}), yet no existing system learns optimal approaches for each user during their own conversation.

In response, AURA addresses these gaps through four integrated contributions. First, it operationalizes response quality as a composite real-time metric that combines length, self-disclosure, emotional intensity, and specificity, enabling automated assessment during conversation rather than relying on post-hoc human evaluation. Next, it employs reinforcement learning to optimize question selection based on observed quality improvements, treating engagement gains as reward signals that guide adaptive strategy choice. In addition, it implements \textit{within-session learning}: the system begins each conversation with priors derived from 96 prior interactions but updates its estimates dynamically as each user responds, discovering personalized strategies within 10--15 exchanges. Finally, it integrates these components into a functioning conversational chatbot that combines the natural-language generation capabilities of large language models with the adaptive decision-making of reinforcement learning, enabling campus-climate surveys that adjust questioning strategies to each respondent’s engagement patterns. Together, these elements establish AURA as the first framework to unify conversational empathy, quantitative quality metrics, and reinforcement learning for real-time adaptive surveying.

The remainder of this paper details AURA’s implementation and evaluation.  
Section~\ref{sec:method} describes the system architecture, including quality-scoring procedures, state representation, action selection, and reinforcement-learning algorithms. Section~\ref{sec:results} presents empirical validation showing that AURA’s adaptive approach yields significant quality improvements over non-adaptive baselines. Finally, Section~\ref{sec:discussion} discusses the broader implications for conversational AI in sensitive organizational-assessment contexts and outlines directions for future research.

\section{Methods}
\label{sec:method}

\subsection{Overview of the AURA Framework}
\label{sec:overview}

AURA (Adaptive Understanding through Reinforcement learning for Assessment) integrates conversational AI with reinforcement learning to enable real-time personalization in campus-climate surveys. Unlike traditional chatbots that follow fixed question sequences, AURA continuously monitors user engagement and adapts its questioning strategy within each conversation based on observed response-quality patterns. This adaptive loop transforms passive data collection into an interactive optimization process, allowing AURA to infer which questioning strategies maximize engagement within a single conversation and ultimately elicit richer, more detailed feedback than static survey instruments. 

\begin{figure}[h]
    \centering
    \includegraphics[width=\textwidth]{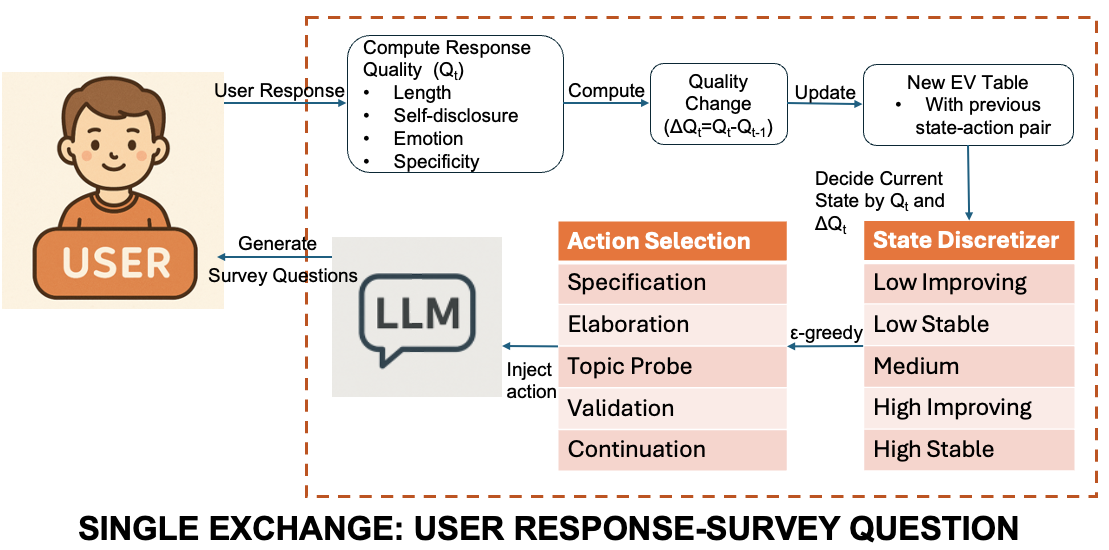}
    \caption{AURA system architecture showing the reinforcement learning cycle 
    within a single conversation exchange. Each user response is scored along 
    four quality dimensions (LSDE), mapped to an engagement state, and used 
    to select the next question type via an $\epsilon$-greedy policy. Observed 
    quality changes update the system's expected-value (EV) estimates, 
    enabling rapid within-session adaptation.}
    \label{fig:AURA_system_architecture}
\end{figure}

Figure~\ref{fig:AURA_system_architecture} illustrates the complete architecture. The system operates as a continuous feedback loop: user responses are assessed for quality, generating reward signals that update action-selection policies in real time. This within-session learning enables the system to discover which questioning strategies work best for each individual participant. 

\begin{figure}[h]
    \centering
    \includegraphics[width=\textwidth]{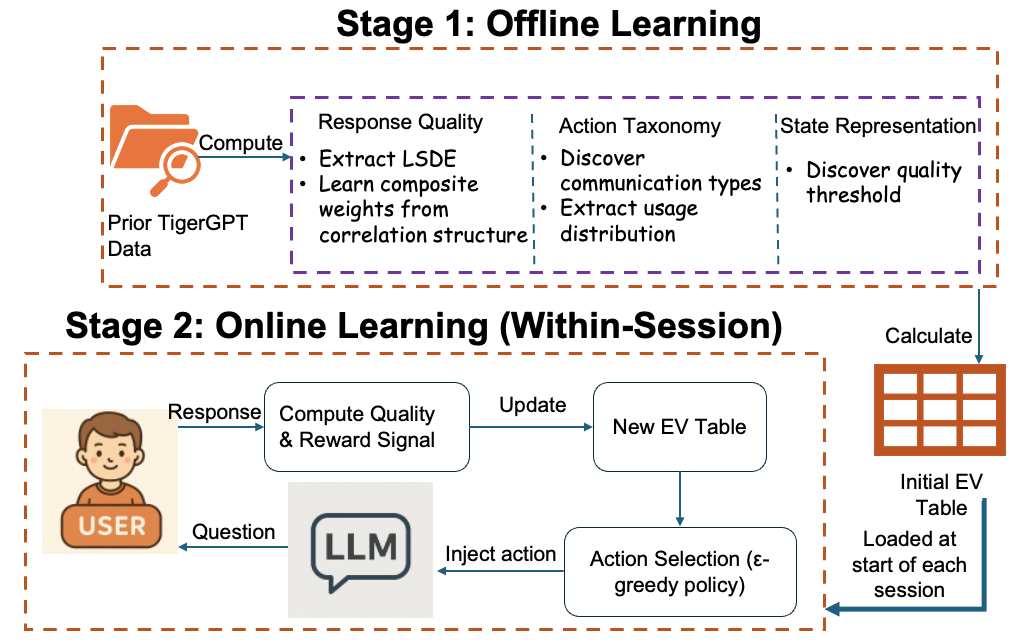}
    \caption{Two-level learning framework. Offline learning (top) extracts patterns from prior conversations to initialize the policy. Online learning (bottom) adapts to individual users within each session through real-time quality feedback, then resets to priors for the next user.}
    \label{fig:AURA_two_level_learning}
\end{figure}

AURA employs a two-level learning strategy that distinguishes it from conventional multi-session reinforcement learning systems (Figure~\ref{fig:AURA_two_level_learning}). At the \textit{offline level}, the system extracts patterns from 96 prior campus-climate conversations to initialize an expected-value (EV) table estimating the quality gain associated with each question type in each engagement state. This offline process establishes population-level priors derived from authentic survey interactions. At the \textit{online level}, during each individual conversation, AURA updates its EV estimates in real time based on the current user's responses, discovering personalized strategies within 10--15 exchanges. Critically, the system resets to priors at the start of each new conversation—no information from previous users is retained—ensuring that adaptive learning remains session-specific and that participant privacy is fully preserved.

This architecture combines the generalizability of population-level patterns with the responsiveness of individual adaptation. By beginning each conversation with empirically grounded priors but adjusting strategy dynamically as user engagement evolves, AURA discovers personalized questioning approaches that static systems cannot achieve. The remainder of this section details the data sources (Section~\ref{sec:data}), quality-assessment methodology (Section~\ref{sec:lsde}), state and action representations (Sections~\ref{sec:actions}--\ref{sec:states}), and reinforcement learning framework (Section~\ref{sec:rl_framework}).

\subsection{Data Collection and Preprocessing}
\label{sec:data}

To initialize AURA's reinforcement-learning policy and establish empirical foundations for engagement measurement, we used campus-climate survey conversations collected from a pilot deployment of TigerGPT \citep{tang2025tigergpt}.  TigerGPT is a semi-adaptive survey chatbot that conducts natural, open-ended dialogues but does not employ reinforcement learning.  The resulting dataset provides authentic examples of engagement in conversational surveys and enables offline estimation of action effectiveness across different engagement states.

\subsubsection{Prior Conversation Dataset}

The dataset comprises 96 campus-climate survey conversations conducted between October 2024 and February 2025 at the University of Missouri.  The chatbot was publicly accessible through a web-based survey link, and all responses were collected anonymously as part of routine feedback activities.  Each conversation included an initial role selection (student, faculty, or staff) followed by optional demographic prompts to personalize dialogue context.  Participants then selected one of five available survey topics—academic issues, financial concerns, work–life balance, campus inclusivity, or a random option—and engaged in multi-exchange dialogues where TigerGPT posed open-ended questions and users provided free-text responses.

\subsubsection{Data Cleaning and Quality Control}

From raw conversation logs totaling 683 exchanges, we applied standard preprocessing procedures:  
(1) removed entries where either the chatbot dialogue or user response was missing, empty, or contained placeholder strings (``nan'', ``N/A''); 
(2) excluded duplicate responses caused by technical errors; and (3) retained only substantive user inputs containing at least one word. 
This yielded a final dataset of 467 valid user responses across 96 conversations. The 31.6\% exclusion rate reflects typical patterns in conversational data collection, where technical issues, participant dropout, or minimal engagement produce incomplete records.

\subsubsection{Dataset Characteristics}

\input{AURA_dataset_summary_table}

Table~\ref{tab:AURA_dataset_summary} summarizes the characteristics of the prior conversation dataset collected from TigerGPT, which serves as the empirical foundation for AURA's reinforcement learning initialization.

The dataset exhibits substantial heterogeneity in conversation length. Single-exchange interactions (29.2\%) occurred when users provided initial feedback but chose not to continue; these responses contribute to cross-sectional quality measurement but cannot inform dynamics of engagement change. Multi-exchange conversations (70.8\%, $N=68$) enable analysis of how response quality evolves across exchanges, producing 371 consecutive exchange pairs used for sequential analysis.

In this context, an \textit{exchange} represents one full exchange between the chatbot and a user. An \textit{exchange pair} links two consecutive user responses via the intervening chatbot question, enabling analysis of how question types affect response quality changes. Since the first response in each conversation has no preceding response for comparison, the 467 total responses yield 371 exchange pairs for sequential analysis ($467 - 96 = 371$).

The longest conversation reached 18 exchanges, though most remained brief (median\,=\,2.5), reflecting typical engagement patterns in voluntary online surveys where respondents balance disclosure with time constraints. This distribution shaped AURA's design requirements: conversations are short enough (mean\,=\,4.9 exchanges) that adaptive systems must learn rapidly, yet long enough (70.8\% with multiple exchanges) to allow repeated state visits and within-session policy refinement.

The remainder of this section details how AURA transforms these prior conversations into an adaptive learning system: Section~\ref{sec:lsde} presents the LSDE quality-scoring methodology applied to all responses, Section~\ref{sec:actions} defines the action space of question types, Section~\ref{sec:states} describes the state representation that discretizes engagement patterns, and Section~\ref{sec:rl_framework} explains the reinforcement learning algorithm that enables within-session personalization.

\subsubsection{Data Usage in AURA Development}

The prior conversation data served three primary functions in AURA's development.

\textbf{Empirical grounding for quality metrics.} 
The 467 user responses provided a reference distribution for establishing normalization parameters and threshold values in the LSDE scoring system (Section~\ref{sec:lsde}). 
For example, the length normalization cap was set at 29~words (the 75th percentile of the observed distribution), ensuring that normalized scores reflect empirical variability rather than arbitrary theoretical bounds.

\textbf{Initial policy estimation.} 
The 371 consecutive exchange pairs enabled the calculation of expected-value (EV) estimates for state–action combinations, capturing population-level patterns in how different question types affect engagement across distinct user states (Section~\ref{sec:rl_framework}). 
These estimates initialize AURA's policy before individual conversations begin, providing evidence-based starting points derived from authentic survey interactions rather than random or uniform priors.

\textbf{Action taxonomy validation.} 
The diversity of question types in the corpus (Section~\ref{sec:actions}) confirmed that the five-action categorization—\textit{specification}, \textit{elaboration}, \textit{topic probe}, \textit{validation}, and \textit{continuation}—captured meaningful functional distinctions in conversational question design. 
Question classification used an LLM-based content analysis (GPT-4o) followed by human validation ($\kappa$\,=\,0.83), establishing both the taxonomic structure and the empirical distribution of action usage in naturalistic campus-climate dialogues.

Importantly, these prior data inform AURA's initialization but do not constrain its behavior during live conversations. Each new user begins with population-level priors, but the system adapts its policy in real time based on individual response patterns, discovering personalized strategies that may diverge substantially from prior averages. The prior corpus thus serves as an empirically grounded prior rather than a fixed reference determining all subsequent decisions.

\subsection{Response Quality Assessment (LSDE)}
\label{sec:lsde}
Effective conversational surveys require real-time measurement of response quality to guide adaptive question selection. We developed a four-dimensional quality metric (Length, Self-disclosure, Emotion, and Specificity; LSDE) that captures distinct aspects of engagement and enables automated assessment during live conversations. This section presents the theoretical foundations for each dimension, describes implementation procedures, and validates the composite scoring approach through empirical analysis of prior campus-climate conversations.

\subsubsection{Foundations and Operationalization}

Survey methodologists emphasize that response quality in open-ended questions is inherently multidimensional: no single indicator fully captures informativeness, engagement, or authenticity \citep{dillman2014internet, tourangeau2000psychology}. Traditional assessment relies on post-hoc human coding, which cannot support real-time adaptive systems. AURA operationalizes four quality dimensions through automated linguistic analysis, enabling exchange-by-exchange measurement during conversation.

\paragraph{Length and Elaboration}

Response length provides a first-order indicator of respondent investment and information content \citep{dillman2014internet, holland2009measuring}. In traditional web surveys, detailed answers ($\approx$50 words) are rated as substantive, but chatbot conversation exchanges are naturally shorter because users elaborate across multiple exchanges.  In our dataset ($N=467$), the median response length was 10 words and the 75th percentile (Q75) was 29 words, providing an empirically grounded scale for this dialogue context.  Length is therefore normalized as
\begin{equation}
L_{\text{norm}} = \min\left(\frac{\text{word\_count}}{29}, 1.0\right)
\end{equation}
so that the top quartile of responses receives full credit for elaboration while maintaining discrimination among briefer exchanges.

\paragraph{Self-Disclosure and Personal Engagement}

First-person pronouns indicate self-focus and willingness to share personal experiences \citep{pennebaker2003linguistic, tausczik2010psychological}.  In survey contexts, self-disclosure enriches qualitative feedback because responses grounded in personal experience (e.g., ``I felt excluded when\dots'') reveal deeper insight than abstract statements (e.g., ``Some students experience exclusion'').  Because our responses are brief (median = 10 words), we use absolute rather than ratio-based pronoun counts to avoid mechanical dependence on response length: as text length increases, ratio scores decline even when disclosure remains constant, producing the strong Length–Disclosure correlation ($r = 0.856$; Table~\ref{tab:AURA_correlation}).  Absolute counts, capped at the 75th-percentile value of three pronouns, measure disclosure directly while retaining discriminative range. Self-disclosure is therefore normalized as

\begin{equation}
D_{\text{norm}} = \min\left(\frac{\text{pronoun\_count}}{3}, 1.0\right)
\end{equation}
where 3 represents the Q75 of our pronoun distribution. Pronouns include \emph{I, me, my, mine, myself, we, us, our, ours, ourselves}, following LIWC's first-person category.

\paragraph{Emotional Expression and Intensity}

Emotional language reflects psychological investment and topic salience.  According to Russell's \citeyearpar{russell1980circumplex} circumplex model of affect, emotion varies along valence (positive–negative) and arousal (intensity).  In surveys, highly affective language—positive or negative—indicates engagement \citep{porter2005mail}.  To measure this automatically, we apply VADER sentiment analysis \citep{hutto2014vader} and use the magnitude of its compound score as a proxy for emotional intensity, independent of polarity.  Although VADER estimates valence rather than arousal directly, research on affective norms shows that extreme valence correlates moderately with arousal ($r \approx 0.40$; \citealt{warriner2013norms}), making the absolute compound value a practical indicator of engagement strength in brief conversational responses. Emotional intensity is therefore normalized as:
\begin{equation}
E_{\text{norm}} = \left| \text{VADER}_\text{compound} \right|
\end{equation}
This transformation yields normalized emotion scores on $[0, 1]$, where 0 indicates affectively neutral language and 1 indicates maximal emotional intensity.

\paragraph{Specificity and Concreteness}

Concrete, detailed responses convey richer and more actionable information than vague generalities.  Episodic memory research shows that specific recollection involves three components—entities (who/what), temporal references (when), and spatial references (where) \citep{levine2002autobiographical}.  Authentic accounts contain more such contextual details than fabricated ones ($d=0.72$; \citealt{newman2003lying}).  Unlike word-level concreteness metrics \citep{brysbaert2014concreteness}, AURA detects the presence of these episodic components as binary indicators of specificity.

Each response is classified by GPT-4o for presence (1) or absence (0) of \textbf{entities}, \textbf{temporal}, and \textbf{spatial} details, following \citeauthor{levine2002autobiographical}'s framework.  Manual inspection of 50 samples confirmed consistent classification accuracy, after which the protocol was applied to the full corpus:
\begin{align}
S_{\text{total}} &= \text{entities} + \text{temporal} + \text{spatial} \in \{0, 1, 2, 3\}, \\
S_{\text{norm}}  &= \frac{S_{\text{total}}}{3}.
\end{align}
Specificity is extremely sparse in our data (88.7\% of responses have $S_{\text{total}} = 0$, Q75 = 0). Although each component is detected binarily, their sum captures meaningful gradations: responses may contain one component (10.1\%), two components (1.1\%), or all three (0.2\%). This count-based approach preserves distinctions among the 11.3\% of responses containing specific detail while remaining robust to the severe sparsity.

\subsubsection{Composite Scoring and Weighting Rationale}
\label{sec:lsde_weighting}

With each dimension normalized to $[0,1]$, we combine them into a composite quality score.  Equal weighting ($0.25$ per dimension) would overrepresent correlated features: Table~\ref{tab:AURA_correlation} shows strong interdependencies among Length, Disclosure, and Emotion ($r=0.52$–$0.86$), whereas Specificity is relatively independent ($\text{mean}\,|r|=0.27$).

\input{AURA_correlation_table}

Unequal weights were therefore assigned to balance three goals: (1) limit redundancy among highly correlated dimensions, (2) emphasize dimensions with broad empirical variation, and (3) retain all four theoretical constructs.  The final composite metric is

\begin{equation}
Q_{\text{composite}} = 0.20 L_{\text{norm}} + 0.20 D_{\text{norm}} + 0.35 E_{\text{norm}} + 0.25 S_{\text{norm}}
\end{equation}

\noindent\textbf{Weighting rationale.}
\begin{itemize}
\item \textbf{Length and Disclosure (0.20 each; 0.40 total):} Highly correlated ($r=0.86$); combined weight limits redundancy while reflecting their joint role in elaboration and engagement.
\item \textbf{Emotion (0.35):} Moderately correlated ($r=0.52$–$0.66$) but theoretically central to engagement intensity and empirically well distributed.
\item \textbf{Specificity (0.25):} Statistically most independent but sparse (88.7 \% zeros).  A moderate weight preserves sensitivity for common responses while rewarding information-rich details.
\end{itemize}

\subsubsection{Empirical Validation}

To evaluate the composite metric, we computed quality scores for all 467 prior responses.  Scores spanned the full $[0,1]$ range (mean = 0.328, SD = 0.276), indicating broad variability and confirming that the metric captures meaningful differences in engagement levels (Table~\ref{tab:AURA_quality_dist}).  

\input{AURA_quality_dist_table}

Nearly half of the responses (47.1\%) fell into the Very Low category, reflecting the engagement challenges typical of campus-climate surveys in which many participants provide only minimal input (e.g., one-word replies).  This distribution demonstrates that the LSDE metric can distinguish low from high engagement with fine granularity, supporting its use as a reward signal for reinforcement learning models that aim to improve response quality through adaptive question selection.

\subsection{Action Space Extraction}
\label{sec:actions}

Effective reinforcement learning for conversational surveys requires a well-defined \textit{action space} representing the questioning strategies available to the system.  Rather than imposing predetermined categories, we extracted action types empirically from prior TigerGPT conversations and mapped them to established communication theory frameworks.  A GPT-4o–based content analysis labeled each survey question into one of five communicative functions—\textit{specification}, \textit{elaboration}, \textit{topic probe}, \textit{validation}, and \textit{continuation}.  The model prompt instructed classification by primary and optional secondary intents, with structured JSON outputs including confidence and brief reasoning.  Human review of a random subset confirmed consistent labeling ($\kappa = 0.83$).  This process yielded a compact yet comprehensive taxonomy suitable for reinforcement-learning–based policy optimization.

\subsubsection{Taxonomy Development: A Theory-Informed, Data-Driven Approach}

Action taxonomy development followed a hybrid methodology balancing empirical observation with theoretical grounding.  Analysis of 467 chatbot questions across 96 TigerGPT conversations (Section~\ref{sec:data}) revealed recurring functional patterns: some prompts requested concrete examples, others sought elaboration, introduced new topics, or offered acknowledgment.  These patterns were consolidated into five communicative categories and validated against established communication frameworks to ensure conceptual coherence \citep{li2016deep, su2021dynamic}.  

This hybrid strategy avoids the limitations of purely data-driven extraction—which can yield uninterpretable clusters—and purely theory-driven design—which may impose artificial categories unrelated to the data.  AURA’s action space is thus both \textbf{empirically grounded} (all categories appear with sufficient frequency in natural dialogues) and \textbf{theoretically justified} (each aligns with established constructs in communication and dialogue management).  

The five-category structure balances interpretability with statistical power.  With 96 conversations producing 371 consecutive exchange pairs, compound multi-intent schemes would create extreme data sparsity.  For instance, allowing two-intent combinations would expand the action space from 5 to 15 categories, and three-intent combinations to 25—yielding only about 15 examples per category on average, with many categories represented by fewer than five instances.  AURA therefore uses primary-intent classifications for reinforcement learning while retaining secondary intents for post-hoc analysis, prioritizing statistical robustness over taxonomic exhaustiveness.

\subsubsection{Action Classification Methodology}

Each chatbot dialogue in the prior corpus was classified by GPT-4o using a structured prompt defining five communicative categories and specifying primary versus secondary intents.  The model identified the main communicative function of each question (primary intent) and any auxiliary functions (secondary intents).  Approximately 65 \% of questions exhibited multiple intents (e.g., “Thank you for sharing. Could you tell me more about that specific example?” combines \textit{validation} and \textit{specification}), illustrating the prevalence of compound strategies in natural dialogue.

Primary intent was defined as the question’s dominant illocutionary force \citep{searle1976classification}—the central action the question seeks to elicit.  Model outputs included structured JSON records with confidence scores and brief reasoning to support quality control and flag ambiguous cases.

To validate classification accuracy, a random sample of 20 questions was human-coded, showing strong agreement between GPT-4o labels and theoretical definitions.  This human-in-the-loop check confirmed that the automated classification captured meaningful functional distinctions rather than superficial lexical cues.

\input{AURA_action-taxonomy}

\subsubsection{Action Type Definitions and Theoretical Grounding}

Table~\ref{tab:AURA_action-taxonomy} presents the five action types with their theoretical foundations and representative examples from the TigerGPT corpus.  Each action corresponds to distinct constructs from Speech Act Theory \citep{searle1976classification} and Communication Accommodation Theory (CAT) \citep{giles1991contexts}, ensuring that AURA’s taxonomy captures functionally meaningful communicative strategies rather than superficial lexical patterns.

\textbf{Specification} and \textbf{elaboration} both operate as directive speech acts but differ in focus.  Specification requests concrete examples or particular cases, using CAT’s convergence strategy to narrow discussion toward actionable details (e.g., transforming “Classes can be challenging” into “Professor Smith’s EECS 280 lectures often run overtime”).  Elaboration maintains the current topic while seeking richer description without requiring specific instantiation (e.g., “Tell me more about your research experience”).  The distinction is functional: specification targets concreteness, elaboration depth.

\textbf{Topic probe} introduces new experiential dimensions, shifting attention to unexplored areas.  These prompts function at the intersection of commissive and directive acts \citep{searle1976classification}, committing the system to expanding the dialogue scope.  They reflect CAT’s divergence strategy \citep{giles1991contexts}, ensuring broad coverage of campus climate dimensions (academic, social, financial, inclusivity).

\textbf{Validation} acknowledges user contributions without requesting new information.  As expressive acts, validation supports rapport and psychological safety through maintenance strategies \citep{giles1991contexts}.  Although rare in prior data ($n=4$, 0.9\%), validation was retained because evidence shows it facilitates deeper disclosure on sensitive topics \citep{wheeless1977measurement, melville2016conducting}.

\textbf{Continuation} invites further input without specifying direction, granting users conversational autonomy.  Also infrequent ($n=2$, 0.4\%), continuation serves as a flow-maintaining strategy that may prove effective for users preferring open-ended engagement \citep{kocaballi2019personalization}.

\input{AURA_action-distribution_table}

\subsubsection{Action Distribution in Prior Data}
\label{sec:histor_action_dist}

Table~\ref{tab:AURA_action-distribution} summarizes the distribution of action types in the prior TigerGPT corpus.  The data show a pronounced imbalance: \textit{specification} dominates (62.3\%), while \textit{validation} (0.9\%) and \textit{continuation} (0.4\%) are rarely used.  This pattern reflects conventional survey design practices that prioritize information extraction over rapport maintenance or open-ended exploration.

From a reinforcement learning perspective, this imbalance presents both a challenge and an opportunity.  The challenge is data sparsity: rare actions have too few examples for reliable expected-value (EV) estimation, creating cold-start conditions for several state–action pairs.  Specifically, eight of twenty-five combinations lack any prior samples (Section~\ref{sec:rl_framework}), requiring AURA to explore them during live interactions.  The opportunity lies in AURA’s capacity to \textbf{correct prior biases}: the dominance of specification likely reflects template-driven defaults rather than optimal engagement strategies.  By adaptively testing underused actions such as \textit{validation} after emotional responses or \textit{topic probe} during stagnation, AURA can discover more effective questioning patterns for individual users.

An $\epsilon$-greedy exploration policy ($\epsilon=0.30$; Section~\ref{sec:rl_framework}) ensures all actions receive exposure despite imbalance.  Roughly 30\% of actions per conversation are randomly selected, allowing even rare strategies to be evaluated.  When these actions improve subsequent response quality, their expected values increase and they are chosen more often for that user; conversely, ineffective actions decline in frequency.  Through this continual rebalancing, AURA learns personalized questioning strategies that static, rule-based systems cannot achieve.

\subsubsection{Multi-Intent Analysis and Primary Action Selection}

About 65 \% of prior questions contained multiple intents, showing that conversational prompts often blend functions.  For instance, “Thank you for sharing that. Could you tell me about a particular instance when this happened?” combines \textit{validation} with \textit{specification}, illustrating how rapport building and information elicitation frequently co-occur.

For reinforcement learning, we label each question by its \textbf{primary intent}—the dominant communicative function.  This choice resolves a key design trade-off: compound actions capture conversational nuance but generate extreme data sparsity.  With five base types, even two-intent combinations expand the action space to 15 categories and three-intent combinations to 25+, leaving fewer than five samples per class across 96 conversations—insufficient for reliable expected-value estimation.

Using primary intent yields a tractable five-action space (25 state–action pairs) while retaining compound information for post-hoc analysis.  Secondary intents were recorded but excluded from training, allowing later evaluation of whether particular combinations (e.g., \textit{specification} + \textit{validation}) consistently enhance engagement.  This design favors \textbf{within-session learning feasibility} over exhaustive taxonomy: AURA must learn effective strategies within 10–15 exchanges per user.  Future extensions could explore hierarchical policies that first select a primary action, then conditionally choose secondary intents as data volume grows.

\subsection{State Representation Design}
\label{sec:states}
Having established AURA's five-action question taxonomy, we now define the state representation that captures user engagement patterns during conversation. Effective reinforcement learning requires discrete states that both reflect meaningful engagement differences and enable sufficient data for policy learning within short conversations. This section describes the state discretization approach, threshold selection rationale, and empirical validation of the five-state framework.

\subsubsection{Design Requirements and Constraints}
State representation in conversational reinforcement learning must balance expressiveness with learnability. AURA faces two critical constraints that shape state design. First, conversations are brief: our prior conversations exhibit a mean of 4.9 exchanges per conversation (median = 2.5), and 70.8\% contain at least two exchanges. Adaptive systems must therefore learn rapidly, discovering effective strategies within the limited exchanges typical of voluntary survey interactions. Second, AURA employs session-specific learning with no cross-user memory retention: the system initializes each conversation with population-level priors but adapts only to the current user, resetting completely before the next interaction. This design preserves privacy and prevents overfitting to specific individuals but requires state spaces that enable meaningful learning from limited within-session experience.

These constraints necessitate discrete rather than continuous state representations. In a typical conversation with continuous states (where each quality score is unique), the system would encounter each state exactly once, precluding any learning from repeated state visits. Discrete states, by contrast, enable state revisits: if a user's engagement remains in the ``medium'' category across multiple exchanges, the system observes which actions succeed or fail in that state, updating its policy accordingly. With 5 discrete states, AURA is designed for conversations of 10-15 exchanges, enabling users to revisit each state 2-3 times on average and providing sufficient feedback for within-session adaptation. Although prior TigerGPT conversations averaged only 4.9 exchanges, AURA's adaptive design aims to sustain longer, more productive dialogues.

\subsubsection{Hybrid State Representation: Quality Level and Trajectory}

We adopt a hybrid approach that combines absolute engagement level with recent trajectory, yielding five states: \texttt{low\_improving}, \texttt{low\_stable}, \texttt{medium}, \texttt{high\_improving}, and \texttt{high\_stable}. Each state is determined by two factors: the current response's composite quality score ($Q_{\text{composite}}$) and the change in quality relative to the previous exchange ($\Delta Q$):

\begin{equation}
\Delta Q_t = Q_{\text{composite}, t} - Q_{\text{composite}, t-1}
\end{equation}

State assignment follows a hierarchical decision rule. First, quality is categorized into three levels based on thresholds derived from the prior data distribution: low ($Q < 0.3$), medium ($0.3 \leq Q < 0.6$), and high ($Q \geq 0.6$). These thresholds were selected to align with meaningful quality distinctions: the low threshold (0.3) approximates the dataset mean (0.328), capturing below-average responses; the high threshold (0.6) exceeds the 75th percentile (Q75 = 0.520), identifying top-quartile engagement.

For low and high quality levels, trajectory information further refines the state. A quality change exceeding $|\Delta Q| > 0.05$ indicates improving engagement, while smaller changes ($|\Delta Q| \leq 0.05$) reflect stable patterns. The 0.05 threshold was calibrated to detect meaningful shifts without excessive sensitivity to measurement noise: it represents approximately one standard deviation of exchange-to-exchange variability in the prior corpus. Responses in the medium quality range are not subdivided by trajectory, as pilot analysis revealed insufficient distinction between improving and stable medium-engagement patterns given the available data.

The complete state assignment function is:

\begin{equation}
\label{eq:state_assignment}
s_t = \begin{cases}
\texttt{low\_improving} & \text{if } Q_t < 0.3 \text{ and } \Delta Q_t > 0.05 \\
\texttt{low\_stable} & \text{if } Q_t < 0.3 \text{ and } \Delta Q_t \leq 0.05 \\
\texttt{medium} & \text{if } 0.3 \leq Q_t < 0.6 \\
\texttt{high\_improving} & \text{if } Q_t \geq 0.6 \text{ and } \Delta Q_t > 0.05 \\
\texttt{high\_stable} & \text{if } Q_t \geq 0.6 \text{ and } \Delta Q_t \leq 0.05
\end{cases}
\end{equation}

For the first exchange of each conversation, where no previous quality score exists, $\Delta Q$ is defined as zero. This assigns the user to either a ``stable'' state (\texttt{low\_stable} or \texttt{high\_stable}) or to \texttt{medium}, depending solely on initial quality level.


\subsubsection{Rationale for Quality-Trajectory Hybrid}
The hybrid design addresses a fundamental question in adaptive surveying: should the system's next action depend only on how engaged the user currently is, or also on whether engagement is increasing or decreasing? Pure quality-level states (e.g., simply ``low,'' ``medium,'' ``high'') ignore dynamics, treating a user whose engagement is declining identically to one whose engagement is improving—even though these patterns may warrant different questioning strategies. Conversely, pure trajectory-based states (e.g., ``improving,'' ``stable,'' ``declining'') disregard absolute engagement levels, potentially recommending the same action for a slightly improving disengaged user and a slightly improving highly engaged user.

The hybrid approach captures both dimensions. For example, \texttt{\seqsplit{low\_improving}} indicates rising quality from a low baseline, suggesting that recent question types may be resonating with the user. By contrast, \texttt{low\_stable} reflects persistent low engagement, indicating that previous strategies have not improved response quality. Similarly, distinguishing \texttt{high\_improving} from \texttt{high\_stable} enables the system to differentiate between users building momentum and those maintaining consistent engagement—patterns that may respond differently to continued questioning. This two-factor representation provides the reinforcement learning algorithm (Section~\ref{sec:rl_framework}) with sufficient information to adapt its strategy based on both current engagement and recent trajectory.

\subsubsection{Empirical Validation and Coverage}
To validate the five-state framework, we categorized all 467 prior responses according to Equation~\ref{eq:state_assignment}. Table~\ref{tab:AURA_state_distribution} presents the resulting distribution. The \texttt{low\_stable} state dominates (46.7\%), reflecting the engagement challenge inherent in voluntary campus-climate surveys: many participants provide minimal responses. The \texttt{medium} state accounts for 27.8\% of responses, representing moderate engagement. High-engagement states (\texttt{high\_improving} and \texttt{high\_stable}) together constitute 16.7\%, confirming that sustained detailed disclosure remains relatively rare in brief survey interactions.

\input{AURA_state_distribution_table}

Critically, all five states are well-represented in the prior data, ensuring sufficient examples for policy initialization (Section~\ref{sec:rl_framework}). Although \texttt{high\_stable} constitutes only 3.9\% of responses—the smallest category—this proportion still represents 18 observations, adequate for establishing baseline patterns while acknowledging that rare states will benefit most from within-session adaptation as the system encounters individual users.

The state categories exhibit meaningful quality separation. Mean composite scores range from 0.071 (\texttt{low\_stable}) to 0.738 (\texttt{high\_stable}), spanning nearly the full measurement range and confirming that the discretization captures substantive engagement differences rather than arbitrary divisions. The trajectory distinction within low and high ranges further refines these categories: \texttt{low\_improving} (mean = 0.184) falls between \texttt{low\_stable} (0.071) and \texttt{medium} (0.454), consistent with users transitioning from disengagement toward moderate participation. This empirical validation confirms that the five-state representation provides both sufficient coverage of engagement patterns and meaningful differentiation for reinforcement learning.

\subsection{Reinforcement Learning Framework}
\label{sec:rl_framework}
Having established AURA's quality-scoring methodology (Section~\ref{sec:lsde}), action taxonomy (Section~\ref{sec:actions}), and state representation (Section~\ref{sec:states}), we present the reinforcement learning algorithm enabling within-session adaptation. AURA employs a two-level architecture (Figure~\ref{fig:AURA_two_level_learning}): offline learning initializes an expected-value (EV) table from 96 prior conversations; online learning adapts the policy in real time based on each user's responses. This section describes both phases and details question generation.

\input{AURA_ev_table}

\subsubsection{Offline Learning: Initial Expected-Value Calculation}
\label{sec:offline}

Before engaging with users, AURA initializes its policy using expected-value estimates derived from 371 prior exchange pairs (Section \ref{sec:data}). Unlike frequency-based approaches that select the most common actions, AURA's EV estimation quantifies each action's effectiveness by combining success probability and improvement magnitude.

\textbf{Expected-Value Formula.} For each state-action pair $(s, a)$:
\begin{equation}
\text{EV}(s, a) = P(\text{improvement} \mid s, a) \times \mathbb{E}[\Delta Q \mid \text{improvement}, s, a]
\end{equation}
where:
\begin{itemize}
\item $P(\text{improvement} \mid s, a)$ is the proportion of instances where action $a$ in state $s$ resulted in quality improvement ($\Delta Q > 0$)
\item $\mathbb{E}[\Delta Q \mid \text{improvement}, s, a]$ is the average quality gain when improvement occurred
\item $\Delta Q = Q_{\text{composite},t+1} - Q_{\text{composite},t}$ is the change in composite quality between consecutive exchanges
\end{itemize}

This formulation captures both the probability and magnitude of quality improvements.

\textbf{Calculation Example.} For the state-action pair (\textit{low\_stable}, \textit{topic probe}):
\begin{itemize}
\item Observations: 20 exchange pairs
\item Improvements: 16 of 20 instances (80\%)
\item Average $\Delta Q$ when improved: $+0.3815$
\item $\text{EV}(\text{low\_stable}, \text{topic\_probe}) = 0.80 \times 0.3815 = 0.305$
\end{itemize}

Table~\ref{tab:AURA_ev_table} presents the complete set of initial EV estimates with sample sizes and confidence indicators. Seventeen of 25 state-action pairs (68\%) have prior data; the remaining 8 pairs (32\%) are initialized to $\text{EV} = 0.0$, representing neutral prior expectation.

\textbf{Coverage and Confidence.} Among pairs with data, five have $n \geq 20$ (reliable estimates), six have $5 \leq n < 20$ (moderate confidence), and six have $n < 5$ (low confidence). The exploration rate ($\epsilon = 0.30$) described in Section \ref{sec:online} ensures that all actions, including those with limited or no prior data, receive trials during individual conversations.

\subsubsection{Online Learning: Within-Session Adaptation}
\label{sec:online}
During each conversation, AURA updates its policy in real time based on 
observed quality changes, enabling rapid personalization. This within-session 
learning combines an $\epsilon$-greedy policy balancing exploration and 
exploitation with an update rule that adjusts EV estimates after each exchange.

\textbf{Reward Signal.} At exchange $t$, the immediate reward is the change in 
composite quality:
\begin{equation}
\label{eq:reward}
r_t = \Delta Q_t = Q_{\text{composite},t} - Q_{\text{composite},t-1}
\end{equation}
For $t=1$, no previous quality exists, so $r_1$ is undefined and no update 
occurs.

\textbf{$\epsilon$-greedy Action Selection.} At exchange $t$, given state $s_t$, 
AURA selects the next action:
\begin{equation}
\label{eq:epsilon_greedy}
a_t = \begin{cases}
\text{random action from } \mathcal{A} & \text{with probability } \epsilon = 0.30 \\
\arg\max_{a \in \mathcal{A}} \text{EV}(s_t, a) & \text{with probability } 1 - \epsilon = 0.70
\end{cases}
\end{equation}
The system explores randomly 30\% of the time, ensuring all actions receive
trials despite prior imbalance (specification 62.3\%; validation 0.9\%, 
continuation 0.4\%). Otherwise, it exploits current knowledge by selecting 
the highest-EV action.

\textbf{EV Update Rule.} After observing reward $r_t$, AURA updates the 
expected value for the state-action pair from the previous exchange:
\begin{equation}
\label{eq:ev_update}
\text{EV}(s_{t-1}, a_{t-1}) \leftarrow \text{EV}(s_{t-1}, a_{t-1}) + \alpha \left[ r_t - \text{EV}(s_{t-1}, a_{t-1}) \right]
\end{equation}
where $\alpha = 0.3$ is the learning rate. This standard temporal-difference 
update \citep{sutton1998reinforcement} occurs at exchange $t$ after observing the 
outcome of action $a_{t-1}$ selected in state $s_{t-1}$ at the previous exchange, 
increasing EV when the observed reward exceeds the prior estimate and 
decreasing it when the action underperforms.

These EV updates are session-specific: each conversation initializes with the prior EV table (Table~\ref{tab:AURA_ev_table}), adapts based on the current user's responses, then discards the modified table at session end, ensuring privacy and individual-focused adaptation.

\textbf{Exchange-by-Exchange Process.} At each exchange $t$, AURA executes the following 
sequence:
\begin{enumerate}
\item Collect and score user response: Compute $Q_t$ via LSDE 
      (Section~\ref{sec:lsde})
\item If $t > 1$: 
      \begin{itemize}
      \item Compute reward: $r_t = Q_t - Q_{t-1}$
      \item Update previous action's value: Apply Equation~\ref{eq:ev_update} 
            to $(s_{t-1}, a_{t-1})$
      \end{itemize}
\item Discretize current state: Map $(Q_t, \Delta Q_t) \rightarrow s_t$ via 
      Section~\ref{sec:states}
\item Select next action: Choose $a_t$ using $\epsilon$-greedy policy 
      (Equation~\ref{eq:epsilon_greedy})
\item Generate question: Produce natural-language prompt 
      (Section~\ref{sec:question_gen})
\end{enumerate}

After 15 exchanges or user termination, the session ends and all within-session 
EV updates are discarded, ensuring the next user begins fresh with priors.


\subsubsection{Question Generation}
\label{sec:question_gen}
The reinforcement learning module selects abstract action types, but users 
interact with natural-language questions. AURA employs GPT-4o-mini to translate 
RL decisions into contextually appropriate prompts.

\textbf{Action-to-Prompt Mapping.} Each action type (Section~\ref{sec:actions}) 
is associated with generation directives that instruct the LLM:
\begin{itemize}
    \item Specification: Request specific examples with contextual details
    \item Elaboration: Request deeper explanation about reasoning or feelings
    \item Topic Probe: Explore related but different aspects of campus life
    \item Validation: Briefly acknowledge contributions (under 20 words)
    \item Continuation: Provide minimal prompt (5-10 words) to sustain flow
\end{itemize}

\textbf{Context Integration.} The prompt includes conversation history 
(previous 2--3 exchanges) and action-specific instructions. For specification 
and elaboration, the LLM deepens the current topic; for topic probe, it 
explores related campus climate dimensions (academic, social, diversity, 
resources, mental health) while maintaining conversational coherence.

\textbf{Generation Parameters.} The LLM uses temperature = 0.7 to generate 
natural variation in phrasing while maintaining contextual appropriateness. 
The system does not employ strict templates; each question is generated 
dynamically based on conversation history and selected action type.

\textbf{Example.} Given action = \textit{topic probe} after discussion about 
Greek life exclusivity, the system generates: \textit{``That's a really 
important perspective. Beyond Greek life, how would you describe the broader 
sense of community and belonging you feel on campus—like in classes, clubs, 
or other social settings?''} The question transitions to related content 
(general social climate) through explicit connection to the previous topic.

By separating strategic decisions (which action) from tactical implementation 
(how to phrase), the system combines RL optimization with LLM fluency.

\section{Evaluation Design}
\label{sec:evaluation_design}

\subsection{Research Questions \& Hypotheses}
This evaluation validates AURA's within-session learning capability and optimizes the exploration parameter ($\epsilon$) for conversational survey applications. Analysis of the prior TigerGPT dataset (Section~\ref{sec:data}) revealed problematic questioning patterns: over-reliance on specification (62.3\%, Table~\ref{tab:AURA_action-distribution}) and severe under-utilization of validation (0.9\%). These patterns reflect common practices in conversational survey design where systems prioritize information extraction over rapport maintenance. We address two research questions:

\textit{RQ1: Does AURA's RL-based adaptation improve response quality compared to prior TigerGPT behavior?} 

We test whether within-session learning produces significant quality improvements ($p < 0.05$, Cohen's $d > 0.50$) over the non-adaptive prior baseline, measured by quality change from initial to final conversation exchanges.

\textit{RQ2: What exploration rate ($\epsilon$) optimizes within-session learning in brief survey conversations?} 

We compare three $\epsilon$-greedy configurations to identify the exploration-exploitation balance that maximizes quality gains within 15-exchange interactions: conservative fixed ($\epsilon=0.15$), moderate fixed ($\epsilon=0.30$), and decaying ($\epsilon: 0.40 \rightarrow 0.05$).

\subsection{Baseline System}
The \textbf{prior distribution baseline} replicates actual TigerGPT behavior from the dataset described in Section~\ref{sec:data}. This semi-adaptive campus climate survey system, deployed October 2024--February 2025, collected the 96 conversations (371 consecutive exchange pairs) used to initialize AURA's expected-value table (Section~\ref{sec:rl_framework}). While TigerGPT employed contextual follow-ups and flexible topic selection, it lacked reinforcement learning and exhibited the action distribution patterns shown in Table~\ref{tab:AURA_action-distribution}: specification 62.3\%, elaboration 23.6\%, topic probe 12.8\%, validation 0.9\%, and continuation 0.4\%.

The baseline implements these empirical action frequencies through weighted random sampling. At each exchange, actions are selected according to the prior probabilities, independent of current engagement state, conversation history, or observed quality patterns. The baseline exhibits \textbf{no state awareness}---all exchanges use identical probabilities regardless of user engagement level---and \textbf{no learning mechanism}---the distribution remains constant throughout and across conversations.

This baseline operationalizes two engagement problems AURA targets: (1) \textbf{specification over-reliance} at 62.3\% risks respondent fatigue through repetitive requests for concrete examples, and (2) \textbf{validation neglect} at 0.9\% undermines rapport-building despite its theoretical importance for sustaining engagement. Improvement over this ecologically valid baseline---representing real survey practice rather than an artificial control---demonstrates that RL adaptation addresses documented methodological challenges in conversational data collection.

\subsection{Evaluation Metrics}

\textbf{Primary metric:} Quality Improvement
\begin{equation}
\Delta Q = Q_{\text{final}} - Q_{\text{initial}}
\end{equation}
where $Q_{\text{initial}}$ and $Q_{\text{final}}$ are composite LSDE scores at Exchange~1 and Exchange~15. This metric isolates adaptive contribution by controlling for baseline engagement, directly measuring AURA's core objective of improving response quality through within-session learning.

\textbf{Secondary metrics:} Final quality ($Q_{\text{final}}$), average quality ($\bar{Q}$), and exchange success rate (proportion of exchanges with $\Delta Q > 0$).

\textbf{Action-level analysis:} Distribution of five action types and action appropriateness (proportion of selections with positive expected value in the prior initialization table, Table~\ref{tab:AURA_ev_table}).

\textbf{Temporal analysis:} Early (Exchange~1--5), mid (6--10), and late (11--15) quality trajectories to characterize adaptation dynamics across conversation phases.

\subsection{Experimental Design}

We employed a between-subjects factorial design with 4 system conditions (3 RL variants + prior baseline) $\times$ 4 simulated user profiles. Each condition comprised $n = 20$ conversations (5 repetitions per profile), yielding 80 total conversations. All RL configurations shared $\alpha = 0.30$ learning rate, 15-exchange conversations, and within-session learning only (complete EV table reset between users to preserve privacy and ensure individual-focused adaptation).

\textbf{RL Exploration Configurations.} Three epsilon variants tested alternative exploration-exploitation hypotheses:
\begin{itemize}
\item \textbf{Configuration~1 ($\epsilon = 0.15$):} Conservative fixed exploration with steady 15\% random action sampling throughout all exchanges. The hypothesis was that low exploration prevents distraction from learned patterns while allowing occasional discovery of alternatives.

\item \textbf{Configuration~2 ($\epsilon = 0.30$):} Moderate fixed exploration with consistent 30\% random sampling. This higher exploration rate was hypothesized to prevent premature convergence to suboptimal policies, particularly important given sparse prior data (only 17 of 25 state-action pairs have examples, Section~\ref{sec:rl_framework}).

\item \textbf{Configuration~3 ($\epsilon: 0.40 \rightarrow 0.05$):} Decaying exploration schedule with high initial exploration (40\%) decreasing linearly to minimal exploration (5\%) by Exchange~15. At each exchange $t$, $\epsilon_t = 0.40 - (0.35 \times t/15)$. The hypothesis was that early exploration discovers effective actions while late exploitation refines learned strategy.
\end{itemize}

\textbf{Simulated User Profiles.} To enable experimental control, reproducibility, and statistical power while maintaining computational feasibility, we employed AI-simulated users rather than human participants. Simulated users were implemented using GPT-4o-mini (temperature~=~0.8) with profile-specific prompts ensuring consistent persona characteristics. Four profiles were selected to represent diverse disciplinary backgrounds and academic levels commonly found in campus populations:
\begin{itemize}
\item \textbf{Biology Senior:} Science-oriented student with moderate engagement style, focused on academic experiences and research opportunities.

\item \textbf{Psychology Junior:} Social science perspective with high emotional expressiveness and interpersonal focus.

\item \textbf{Computer Science Sophomore:} Technical background with analytical communication style and emphasis on academic rigor.

\item \textbf{English Senior:} Humanities orientation with elaborative responses and attention to campus culture and community.
\end{itemize}

This disciplinary diversity (STEM, social science, humanities) combined with variation in academic seniority (sophomore, junior, senior) ensured that evaluation captured a range of communication styles and engagement patterns typical of university survey populations. Profile characteristics remained constant across all system conditions, ensuring fair comparison and isolating the effect of questioning strategy from individual differences.

The use of simulated users enabled three critical experimental advantages: (1)~precise control, with identical profiles tested across all conditions; (2)~perfect reproducibility, with conversations deterministically reconstructable from logs; and (3)~adequate statistical power ($> 0.80$ for $d \geq 0.65$ at $\alpha = 0.05$, given $n = 20$ per condition). While simulated users cannot capture the full complexity of human behavior, they provide a controlled environment for validating within-session learning mechanisms before deployment with actual survey participants.

\textbf{Statistical Analysis.} We employed independent samples $t$-tests comparing each RL configuration to the prior baseline. Three planned comparisons were conducted with uncorrected $p$-values reported ($\alpha = 0.05$, two-tailed). Cohen's $d$ effect sizes were computed for all comparisons, with interpretative guidelines: small ($d = 0.20$--$0.49$), medium ($d = 0.50$--$0.79$), large ($d \geq 0.80$).

\section{Results}
\label{sec:results}

\subsection{Overall Performance Comparison}
\label{sec:overall_result}

\input{AURA_overall_performance_table}

Table~\ref{tab:AURA_overall_results} presents primary results comparing three RL configurations to the prior baseline. Configuration~2 ($\epsilon = 0.30$) significantly outperformed the baseline on quality improvement ($t(38) = 2.088$, $p = 0.044$, $d = 0.660$), demonstrating that moderate fixed exploration enables effective within-session adaptation. Configuration~2 also achieved substantially higher final quality scores ($Q_{\text{final}} = 0.582$ vs. baseline $0.465$, a difference of $+0.117$). Configuration~1 ($\epsilon = 0.15$) showed marginal improvement ($p = 0.078$, $d = 0.573$), while Configuration~3 (decaying epsilon) failed to reach significance ($p = 0.345$, $d = 0.302$).

The results provide clear answers to our research questions. For \textbf{RQ1}, RL-based adaptation yields significant quality improvements over non-adaptive baseline behavior when properly configured. The best-performing system (Configuration~2) achieved a quality improvement advantage of $+0.076$ relative to baseline (effect size $d = 0.660$, medium effect), and final quality exceeded baseline by $+0.117$. These effects demonstrate that within-session learning successfully addresses the engagement problems documented in the prior data.

For \textbf{RQ2}, moderate fixed exploration ($\epsilon = 0.30$) proved optimal for brief survey conversations. Configuration~1's conservative exploration ($\epsilon = 0.15$) approached but did not achieve significance, suggesting insufficient sampling leads to premature convergence (examined further in Section~\ref{sec:phase_analysis}). Configuration~3's decaying schedule ($\epsilon: 0.40 \rightarrow 0.05$) produced the weakest performance, indicating that high early exploration introduces excessive variability that disrupts learning before the system can benefit from increased exploitation.

Notably, the prior baseline exhibited minimal quality change ($\Delta Q = -0.006$), indicating that the non-adaptive fixed-probability strategy produced no measurable improvement in engagement over 15 exchanges. In contrast, all three RL configurations achieved positive quality improvement, with Configuration~2 achieving a final quality advantage of $+0.117$ over baseline. This demonstrates that adaptive learning not only improves over initial engagement levels but also maintains higher engagement throughout the conversation compared to static questioning strategies.

\subsection{Temporal Adaptation Dynamics}
\label{sec:phase_analysis}

Section~\ref{sec:overall_result} established that Configuration~2 ($\epsilon = 0.30$) significantly outperformed the prior baseline on overall quality improvement ($p = 0.044$, $d = 0.660$), while Configuration~1 showed only marginal effects and Configuration~3 failed to reach significance. To understand the mechanisms underlying these results---specifically, how within-session learning unfolds and why moderate exploration proves superior---we analyzed quality improvement in three temporal phases: early (Exchanges~1--5), mid (Exchanges~6--10), and late (Exchanges~11--15). This decomposition reveals critical insights into the progression of RL adaptation and identifies the failure modes that distinguish successful from unsuccessful exploration strategies. Table~\ref{tab:AURA_phase_results} presents phase-specific comparisons for all conditions.

\input{AURA_phase_results_table}

\textbf{Early phase} (Exchanges~1--5) showed no significant differences in within-phase quality improvement across conditions. All systems, including the prior baseline, achieved modest positive quality changes ($\Delta Q \approx +0.06$), suggesting that initial engagement benefits from novelty effects and general conversational scaffolding regardless of adaptive strategy. At this stage, RL systems have accumulated insufficient exchange data to differentiate effective policies through learning, relying primarily on population-level priors from the prior initialization (Section~\ref{sec:offline}). However, it is important to note that Configuration~2 achieved significantly higher absolute quality levels during this early phase (mean quality: 0.542 vs. baseline 0.493, $p = 0.006$, $d = 0.392$), indicating that moderate exploration enabled more effective initial question selection even before substantial within-session learning occurred. This early advantage provided a foundation for cumulative quality gains throughout the conversation.

\textbf{Mid phase} (Exchanges~6--10) exhibited high variability with no statistically significant performance differentiation across conditions. Quality improvements were near zero across all systems, with Configuration~2 showing slight within-phase degradation ($\Delta Q = -0.016$), Configuration~1 near zero ($\Delta Q = +0.004$), and Configuration~3 showing modest gains ($\Delta Q = +0.041$). This phase represents a critical transition period where learning mechanisms begin discovering engagement patterns but have not yet converged to stable policies. The apparent decline in Configuration~2 reflects normal fluctuation in quality trajectories rather than systematic failure---crucially, Configuration~2 maintained the highest absolute quality levels throughout this phase (mean quality: 0.559 vs. baseline 0.519, $p = 0.013$, $d = 0.378$), demonstrating sustained effectiveness despite temporary within-phase variance. Notably, Configuration~1 displayed the highest variability ($\pm 0.193$), suggesting unstable policy updates, while Configuration~2 maintained the most consistent performance ($\pm 0.117$), indicating more reliable learning dynamics.

\textbf{Late phase} (Exchanges~11--15) revealed critical differences in adaptation sustainability and exposed a fundamental failure mode in conservative exploration strategies. Configuration~1 ($\epsilon = 0.15$) exhibited significant \emph{negative} quality change ($\Delta Q = -0.116$) compared to prior baseline ($t(38) = -2.357$, $p = 0.025$, $d = -0.862$), indicating that conservative exploration led to premature convergence to a suboptimal policy that actively degraded engagement in later exchanges. This late-phase collapse represents a critical failure of insufficient exploration: the system locked onto learned patterns too quickly, preventing discovery of more effective strategies as conversation context evolved. Paradoxically, despite this degradation, Configuration~1 achieved the highest absolute quality in this phase (0.562) due to strong cumulative gains from earlier exchanges---highlighting that the degradation represents a \emph{failure to sustain adaptation} rather than overall poor performance. The system performed well early but could not maintain its trajectory.

In contrast, Configuration~2 ($\epsilon = 0.30$) maintained positive quality improvement ($\Delta Q = +0.043$) throughout the late phase, outperforming both baseline and Configuration~1 in trajectory sustainability. While this late-phase improvement did not achieve statistical significance ($p = 0.544$), it demonstrates consistent positive adaptation without the collapse observed in Configuration~1. This finding explains why Configuration~2 achieved overall significance in Section~\ref{sec:overall_result}: sustained late-phase performance and stable adaptation throughout the conversation drove the cumulative quality gains that produced the significant overall effect ($\Delta Q = +0.076$ relative to baseline, $p = 0.044$). Configuration~3 ($\epsilon: 0.40 \rightarrow 0.05$) showed slight negative change ($\Delta Q = -0.016$), suggesting that excessive early exploration disrupted coherent policy formation, and subsequent low exploration prevented recovery.

The phase-specific analysis demonstrates that within-session learning in brief conversational contexts requires sufficient exploration to avoid premature policy convergence while maintaining enough exploitation to capitalize on discovered patterns. Configuration~1's conservative 15\% exploration rate proved inadequate for the data-sparse environment (17 of 25 state-action pairs with prior examples, Table~\ref{tab:AURA_ev_table} in Section~\ref{sec:offline}), causing the system to lock onto suboptimal strategies that failed in later exchanges. With only 2--3 exploration trials across 15 exchanges, the system had insufficient sampling diversity to discover effective adaptations for evolving conversation states. Configuration~2's 30\% exploration maintained enough sampling diversity (4--5 exploration trials) to discover effective adaptations while still exploiting learned patterns, exhibiting the most consistent performance across all phases (standard deviations: 0.101--0.131 compared to Configuration~1's 0.143--0.193). This balanced approach enabled the system to avoid both premature convergence (Configuration~1's failure mode) and excessive randomness (Configuration~3's disruption from 40\% initial exploration).

Configuration~3's decaying schedule began with excessive exploration (40\%), introducing noise that disrupted early learning and prevented coherent policy development. By the time exploration decreased to facilitate exploitation, the system had accumulated noisy value estimates that hindered effective decision-making. This finding challenges the common assumption in RL that exploration should decay over time---in brief, high-stakes interactions where every exchange matters, moderate fixed exploration outperforms adaptive schedules. The fixed moderate rate allows the system to continuously validate learned patterns while remaining open to discovering better strategies as conversation context shifts, a critical capability for adaptive survey dialogues where user engagement states evolve unpredictably within single sessions.

These temporal dynamics reveal a fundamental design principle for within-session RL in conversational systems: moderate fixed exploration ($\epsilon = 0.30$) outperforms both conservative ($\epsilon = 0.15$) and decay-based schedules ($\epsilon: 0.40 \rightarrow 0.05$) in brief interactions. The failure modes differ meaningfully: insufficient exploration causes premature convergence and late-phase collapse, while excessive early exploration disrupts coherent policy formation. For survey applications and other brief goal-oriented dialogues, maintaining steady moderate exploration throughout the interaction proves more effective than traditional exploration-decay strategies optimized for long-horizon reinforcement learning tasks. These findings provide clear mechanistic support for the conclusion in Section~\ref{sec:overall_result} that Configuration~2 optimally addresses RQ2, establishing $\epsilon = 0.30$ as the preferred exploration rate for within-session conversational adaptation.

\subsection{Action Distribution Analysis}
\label{sec:action_analysis}

A key question for adaptive conversational systems is whether learning produces behaviorally meaningful changes in questioning strategies. Table~\ref{tab:AURA_action_dist} presents action distributions across all four conditions, revealing the specific behavioral adaptations that underlie the quality improvements demonstrated in Sections~\ref{sec:overall_result} and \ref{sec:phase_analysis}.

\input{AURA_action_dist_result_table}

The prior baseline accurately reproduced TigerGPT's empirical distribution (specification 58.5\% vs. 62.3\% in Table~\ref{tab:AURA_action-distribution}, within sampling variability), validating the baseline implementation. All three RL configurations produced dramatic shifts in action selection, indicating that the system learned qualitatively different questioning strategies rather than minor parameter adjustments:

\textbf{Specification reduction.} The most striking behavioral change was a 62\% reduction in specification usage (58.5\% $\rightarrow$ 22.2\% for Configuration~2), demonstrating that the RL system learned to avoid over-reliance on requesting concrete examples---the primary engagement problem identified in prior data (Section \ref{sec:histor_action_dist}). The consistent reduction across all three RL configurations (57--62\% relative decrease) suggests this shift reflects discovered effectiveness patterns rather than random exploration artifacts. This dramatic rebalancing addresses the prior specification over-reliance (62.3\% in Table~\ref{tab:AURA_action-distribution}) that risked respondent fatigue through repetitive questioning.

\textbf{Validation increase.} Configuration~2 increased validation usage nearly 10-fold (0.8\% $\rightarrow$ 7.9\%), directly addressing the prior under-utilization problem documented in Section~\ref{sec:histor_action_dist}. This finding is particularly notable because validation had minimal prior examples ($n = 4$ total instances, Table~\ref{tab:AURA_action-distribution}), creating a cold-start problem where initial EV estimates were highly uncertain (EV = 0.348 with only 4 observations in the low\_stable state, Table~\ref{tab:AURA_ev_table}). The system's discovery that validation improves engagement---despite sparse prior data and requiring exploration-driven sampling---demonstrates effective learning from limited examples. Configuration~2's higher validation rate (7.9\% vs. 2.9--3.8\% for other RL systems) likely contributes to its superior quality performance, as validation provides rapport-building acknowledgment without demanding additional user effort.

\textbf{Elaboration shift.} All RL configurations substantially increased elaboration usage (27.4\% $\rightarrow$ 48--53\%), suggesting the system discovered that asking users to expand on previous responses yields higher quality than repeatedly requesting new specific examples. This shift aligns with conversational scaffolding principles where building on established topics maintains coherence and reduces cognitive load. However, the magnitude of this shift (approximately doubling elaboration frequency) may indicate over-correction, where the system over-exploits elaboration at the expense of other potentially effective actions. The similar elaboration rates across all three RL configurations (48--53\%) suggest this pattern emerges from prior EV estimates (Table~\ref{tab:AURA_ev_table}) rather than within-session learning, as all systems initialized with the same priors favoring elaboration in multiple states.

\textbf{Continuation adoption.} RL systems increased continuation usage from near-zero (0.4\%) to moderate levels (6--10\%), indicating the system learned to employ minimal follow-up prompts when users demonstrate sustained engagement. This represents discovery of a questioning strategy with virtually no prior precedent ($n = 2$ prior instances, Table~\ref{tab:AURA_action-distribution}), requiring pure exploration-driven learning. Configuration~1 showed the highest continuation usage (10.2\%), possibly compensating for its insufficient exploration in other action dimensions---the system may have defaulted to continuation when uncertain about more strategic alternatives. Configuration~2's more moderate continuation rate (8.8\%) suggests better-calibrated action selection.

\textbf{Topic probe stability.} Configuration~2 maintained topic probe frequency near prior levels (13.0\% vs. 12.9\% baseline), while Configuration~1 decreased topic probe usage (8.2\%). This stability in Configuration~2 demonstrates balanced exploration across action types rather than over-specialization. Topic probe serves a distinct function---introducing new conversation dimensions---that complements rather than competes with elaboration and specification. Configuration~2's preservation of topic probe usage while dramatically shifting other actions suggests more effective policy learning.

Configuration~2's action distribution most effectively balanced these shifts: it achieved the highest validation usage (7.9\%), lowest specification usage (22.2\%), maintained topic probe frequency near prior levels (13.0\% vs. 12.9\%), and showed moderate continuation and elaboration rates. This balanced portfolio of actions likely contributes to Configuration~2's superior overall performance (Section~\ref{sec:overall_result}), avoiding both the specification trap of the prior baseline and potential over-specialization on elaboration observed in Configuration~1 (53.3\%). The distribution demonstrates that effective within-session learning requires not just discovering individual effective actions, but learning when to deploy each action appropriately---a capability that Configuration~2's 30\% exploration rate enabled through sufficient sampling diversity across conversation contexts.

\subsection{Summary and Interpretation}
\label{sec:result_summary}

The evaluation results provide strong evidence that within-session reinforcement learning improves conversational survey quality compared to non-adaptive baseline behavior. Configuration~2 ($\epsilon = 0.30$) achieved statistically significant improvements in both overall quality change ($p = 0.044$, $d = 0.660$) and final engagement levels ($p = 0.005$, $d = 0.949$), validating AURA's core adaptive learning mechanism.

Three key findings emerge from the analysis:

\textbf{1. Exploration rate critically affects adaptation sustainability.} While all RL configurations learned to shift action distributions away from problematic prior patterns, only moderate exploration ($\epsilon = 0.30$) maintained performance gains across all conversation phases. Conservative exploration ($\epsilon = 0.15$) led to late-phase degradation through premature convergence, while decaying exploration schedules introduced excessive early variability that prevented effective learning.

\textbf{2. Behavioral changes reflect discovered effectiveness patterns.} The dramatic reduction in specification usage (63\%), substantial increase in validation (10-fold), and adoption of previously rare continuation actions demonstrate that the system discovered questioning strategies more effective than prior practice. These shifts cannot be attributed to random exploration given their consistency across RL configurations and their alignment with communication theory principles (validation supports rapport, excessive specification causes fatigue).

\textbf{3. Within-session learning succeeds despite sparse prior data.} AURA's policy initialization relied on only 96 prior conversations with substantial action imbalance (17 of 25 state-action pairs had examples). Despite these limitations, the system achieved significant quality improvements within 15-exchange interactions, suggesting that moderate exploration combined with rapid online learning ($\alpha = 0.30$) enables effective adaptation from limited priors.

These results establish Configuration~2 ($\epsilon = 0.30$, $\alpha = 0.30$) as the recommended parameter setting for deploying AURA in conversational survey applications. The system addresses documented engagement problems from prior practice while maintaining computational feasibility and privacy preservation through session-specific learning.

\section{Discussion and Future Work}
\label{sec:discussion}

The findings demonstrate that reinforcement learning can meaningfully enhance conversational survey systems by enabling real-time adaptation within individual sessions.  AURA’s reinforcement-driven questioning strategy achieved consistent quality improvements compared with non-adaptive baselines, confirming that even short survey interactions benefit from dynamic policy updates guided by engagement feedback.  The observed +0.12 mean gain and medium-to-large effect size ($d=0.66$) show that within-session learning is both statistically and practically significant.

These gains are not random fluctuations but reflect the discovery of more effective conversational behaviors.  The sharp reduction in specification prompts, coupled with a tenfold increase in validation and moderate adoption of continuation prompts, indicates that the model internalized strategies consistent with interpersonal-communication theory—balancing information seeking with rapport building to sustain participation.  In doing so, AURA links computational learning with psychological principles of empathy and self-disclosure, translating them into measurable conversational patterns.  Methodologically, it demonstrates how linguistic quality metrics (LSDE) can serve as real-time reinforcement signals, allowing chatbots to approximate human interviewer responsiveness while remaining scalable and transparent.

Several limitations merit attention.  The evaluation used AI-simulated participants to ensure reproducibility and controlled exploration testing; future human-subject studies will be required to confirm generalizability. The prior corpus used for initialization was relatively small and imbalanced, which constrains the richness of early state–action estimates.  Moreover, the study focused on linguistic quality metrics rather than full user-experience factors such as trust or comfort.  Finally, while session-specific learning preserves privacy, it restricts cross-user generalization, highlighting the need for privacy-preserving or federated learning extensions.

Future work will address these limitations through large-scale human evaluations across institutions, expanded datasets, and hybrid online–offline learning schemes that blend personalization with shared optimization.  Broader deployment in domains such as education, mental-health screening, and organizational assessment will test AURA’s adaptability under diverse conversational goals.  Overall, AURA demonstrates that reinforcement learning can endow survey chatbots with genuine adaptivity—learning from each exchange while preserving user autonomy and privacy.  Although validated here through campus-climate assessment, the framework offers a generalizable foundation for adaptive conversational surveys across domains such as education, healthcare, and organizational analytics.

\section*{Acknowledgment}
During the preparation of this paper, the authors used ChatGPT in order to improve the spelling, grammar, and overall readability. After using this tool, the authors reviewed and edited the content as needed and take full responsibility for the content of the publication.

\bibliographystyle{elsarticle-harv} 
\bibliography{cas-refs}

\end{document}

%% file: AURA_dataset_summary_table.tex
\begin{table}[!ht]
\centering
\caption{Summary Statistics of the Prior Conversation Dataset.}
\label{tab:AURA_dataset_summary}
\begin{tabular}{
    @{\hspace{3mm}}l
    @{\hspace{6mm}}r
}
\toprule
\textbf{Category} & \textbf{Value} \\
\midrule
\multicolumn{2}{l}{\textit{Conversation-level}} \\[1pt]
\quad Total conversations & 96 \\
\quad Mean exchanges per conversation & 4.9\,$\pm$\,5.5 \\
\quad Median exchanges per conversation & 2.5 \\
\quad Range of exchanges & 1–18 \\
\quad Single-exchange conversations & 28\,(29.2\%) \\
\quad Multi-exchange conversations (2+ exchanges) & 68\,(70.8\%) \\
\addlinespace[4pt]
\multicolumn{2}{l}{\textit{Response-level}} \\[1pt]
\quad Total valid responses & 467 \\
\quad Consecutive exchange pairs (for sequential analysis) & 371 \\
\quad Mean response length (words) & 18.3\,$\pm$\,22.2 \\
\quad Median response length (words) & 10.0 \\
\bottomrule
\end{tabular}
\end{table}

%% file: AURA_correlation_table.tex
\begin{table}[htbp]
\centering
\caption{Correlation Matrix of Normalized LSDE Dimensions ($N = 467$).}
\label{tab:AURA_correlation}
\begin{tabular}{lcccc}
\toprule
            & Length & Specificity & Disclosure & Emotion \\
\midrule
Length      & 1.000  & 0.290       & 0.856      & 0.664   \\
Specificity & 0.290  & 1.000       & 0.248      & 0.259   \\
Disclosure  & 0.856  & 0.248       & 1.000      & 0.523   \\
Emotion     & 0.664  & 0.259       & 0.523      & 1.000   \\
\bottomrule
\end{tabular}
\end{table}

%% file: AURA_quality_dist_table.tex
\begin{table}[htbp]
\centering
\caption{Distribution of Composite Quality Scores ($N = 467$).}
\label{tab:AURA_quality_dist}
\begin{tabular}{lcc}
\toprule
Category  & Score Range & $n$ (\%) \\
\midrule
Very Low  & 0.0--0.2    & 220 (47.1\%) \\
Low       & 0.2--0.4    & 80 (17.1\%)  \\
Medium    & 0.4--0.6    & 90 (19.3\%)  \\
High      & 0.6--0.8    & 70 (15.0\%)  \\
Very High & 0.8--1.0    & 7 (1.5\%)    \\
\bottomrule
\end{tabular}
\end{table}

%% file: AURA_action-taxonomy.tex
\begin{table}[htbp]
\centering
\small
\begin{threeparttable}
\caption{Action Type Taxonomy, Theoretical Grounding, and Prior Examples}
\label{tab:AURA_action-taxonomy}
\begin{tabularx}{\textwidth}{p{2cm}p{2cm}p{2.5cm}X}
\toprule
\textbf{Action Type} & \textbf{Speech Act Function} & \textbf{CAT Strategy} & \textbf{Example} \\
\midrule
specification & Directive & Convergence (narrowing focus) & ``Could you share how satisfied you are with your academic program and why?'' \\
\addlinespace
elaboration & Directive & Maintenance (sustaining topic) & ``Could you share more about the challenges you've faced in Greek life?'' \\
\addlinespace
topic probe & Commissive & Divergence (expanding scope) & ``How comfortable do you feel on campus, and have you experienced moments of inclusion or exclusion?'' \\
\addlinespace
validation & Expressive & Maintenance (rapport building) & ``Thank you for sharing that. Your perspective is valuable and helps us understand student experiences.'' \\
\addlinespace
continuation & Directive & Maintenance (flow sustaining) & ``Is there anything else you'd like to share about your experiences at MU?'' \\
\bottomrule
\end{tabularx}
\begin{tablenotes}[flushleft]
\small
\item Note: CAT = Communication Accommodation Theory \citep{giles1991contexts}. 
\item Speech act categories follow \citet{searle1976classification}'s taxonomy of illocutionary forces.
\end{tablenotes}
\end{threeparttable}
\end{table}

%% file: AURA_action-distribution_table.tex
\begin{table}[htbp]
\centering
\normalsize
\begin{threeparttable}
\caption{Action Type Distribution in Prior Corpus ($N=467$ conversation exchanges)}
\label{tab:AURA_action-distribution}
\begin{tabularx}{\textwidth}{@{}lrrX@{}}
\toprule
\textbf{Action Type} & \textbf{Count} & \textbf{Percentage (\%)} & \textbf{Characteristics} \\
\midrule
Specification & 291 & 62.3 & Requests concrete examples \\
Elaboration & 110 & 23.6 & Asks for expanded detail \\
Topic probe & 60 & 12.8 & Introduces new dimensions \\
Validation & 4 & 0.9 & Acknowledges contributions \\
Continuation & 2 & 0.4 & General follow-up \\
\midrule
\textbf{Total} & \textbf{467} & \textbf{100} & \\
\bottomrule
\end{tabularx}
\begin{tablenotes}[flushleft]
\normalsize
\item \textit{Note:} Distribution reflects prior TigerGPT behavior. AURA's RL framework enables adaptive rebalancing for individual users.
\end{tablenotes}
\end{threeparttable}
\end{table}

%% file: AURA_state_distribution_table.tex
\begin{table}[h]
\centering
\caption{State Distribution in Prior Conversation Data (N = 467 responses).}
\label{tab:AURA_state_distribution}
\begin{tabular}{lrrr}
\toprule
State & Count & Percentage (\%) & Mean Quality Score \\
\midrule
\texttt{low\_stable} & 218 & 46.7 & 0.071 \\
\texttt{medium} & 130 & 27.8 & 0.454 \\
\texttt{high\_improving} & 60 & 12.8 & 0.709 \\
\texttt{low\_improving} & 41 & 8.8 & 0.184 \\
\texttt{high\_stable} & 18 & 3.9 & 0.738 \\
\midrule
Total & 467 & 100.0 & 0.328 \\
\bottomrule
\end{tabular}
\end{table}

%% file: AURA_ev_table.tex
\begin{table}[htbp]
\centering
\begin{sideways}
\begin{minipage}{\textheight}
\centering
\caption{Initial Expected-Value Estimates from Prior Data (N = 371 exchange pairs)}
\label{tab:AURA_ev_table}
\normalsize
\begin{tabular}{lrrrrr}
\toprule
\textbf{State} & \textbf{Specification} & \textbf{Elaboration} & \textbf{Topic Probe} & \textbf{Validation} & \textbf{Continuation} \\
\midrule
low\_improving & 
    \begin{tabular}[t]{@{}r@{}}0.058\\(n=15, M)\end{tabular} & 
    \begin{tabular}[t]{@{}r@{}}0.047\\(n=9, M)\end{tabular} & 
    \begin{tabular}[t]{@{}r@{}}0.032\\(n=3, L$^\dagger$)\end{tabular} & 
    \begin{tabular}[t]{@{}r@{}}0.000\\(n=0, N$^\ddagger$)\end{tabular} & 
    \begin{tabular}[t]{@{}r@{}}0.000\\(n=0, N$^\ddagger$)\end{tabular} \\
\midrule
low\_stable & 
    \begin{tabular}[t]{@{}r@{}}\textbf{0.288}\\\textbf{(n=112, R)}\end{tabular} & 
    \begin{tabular}[t]{@{}r@{}}\textbf{0.170}\\\textbf{(n=27, R)}\end{tabular} & 
    \begin{tabular}[t]{@{}r@{}}\textbf{0.305}\\\textbf{(n=20, R)}\end{tabular} & 
    \begin{tabular}[t]{@{}r@{}}0.348\\(n=4, L$^\dagger$)\end{tabular} & 
    \begin{tabular}[t]{@{}r@{}}0.476\\(n=1, L$^\dagger$)\end{tabular} \\
\midrule
medium & 
    \begin{tabular}[t]{@{}r@{}}\textbf{0.071}\\\textbf{(n=66, R)}\end{tabular} & 
    \begin{tabular}[t]{@{}r@{}}\textbf{0.073}\\\textbf{(n=28, R)}\end{tabular} & 
    \begin{tabular}[t]{@{}r@{}}0.039\\(n=22, M)\end{tabular} & 
    \begin{tabular}[t]{@{}r@{}}0.000\\(n=0, N$^\ddagger$)\end{tabular} & 
    \begin{tabular}[t]{@{}r@{}}0.000\\(n=0, N$^\ddagger$)\end{tabular} \\
\midrule
high\_improving & 
    \begin{tabular}[t]{@{}r@{}}0.004\\(n=33, M)\end{tabular} & 
    \begin{tabular}[t]{@{}r@{}}0.020\\(n=14, M)\end{tabular} & 
    \begin{tabular}[t]{@{}r@{}}0.000\\(n=4, L$^\dagger$)\end{tabular} & 
    \begin{tabular}[t]{@{}r@{}}0.000\\(n=0, N$^\ddagger$)\end{tabular} & 
    \begin{tabular}[t]{@{}r@{}}0.000\\(n=0, N$^\ddagger$)\end{tabular} \\
\midrule
high\_stable & 
    \begin{tabular}[t]{@{}r@{}}0.040\\(n=9, M)\end{tabular} & 
    \begin{tabular}[t]{@{}r@{}}0.083\\(n=1, L$^\dagger$)\end{tabular} & 
    \begin{tabular}[t]{@{}r@{}}0.028\\(n=3, L$^\dagger$)\end{tabular} & 
    \begin{tabular}[t]{@{}r@{}}0.000\\(n=0, N$^\ddagger$)\end{tabular} & 
    \begin{tabular}[t]{@{}r@{}}0.000\\(n=0, N$^\ddagger$)\end{tabular} \\
\bottomrule
\end{tabular}
\vspace{2mm}

\normalsize
\textit{Note:} Each cell shows EV value with sample size and confidence category in parentheses. Confidence: \textbf{R}eliable ($n \geq 20$), M(oderate) ($5 \leq n < 20$), L(ow)$^\dagger$ ($1 \leq n < 5$), N(one)$^\ddagger$ ($n = 0$, cold-start). EV represents expected quality improvement.
\end{minipage}
\end{sideways}
\end{table}

%% file: AURA_overall_performance_table.tex
\begin{sidewaystable}[htbp]
\centering
\begin{threeparttable}
\caption{Overall Performance Comparison: RL Configurations vs. Prior Baseline.}
\label{tab:AURA_overall_results}

\begin{tabular}{lcccccc}
\toprule
\textbf{Condition} & \textbf{$\Delta Q$} & \textbf{vs. Baseline} & \textbf{$p$-value} & \textbf{Cohen's $d$} & \textbf{$\bar{Q}$} & \textbf{$Q_{\text{final}}$} \\
\hline
Prior Baseline & $-0.006 \pm 0.118$ & --- & --- & --- & $0.513 \pm 0.127$ & $0.465 \pm 0.137$ \\
Config 1 ($\epsilon=0.15$) & $+0.087 \pm 0.190$ & $+0.093$ & $0.078$ & $+0.573$ & $0.535 \pm 0.124$ & $0.538 \pm 0.132$ \\
Config 2 ($\epsilon=0.30$) & $+0.070 \pm 0.106$ & $+0.076$ & $0.044^*$ & $+0.660$ & $0.553 \pm 0.102$ & $0.582 \pm 0.099$ \\
Config 3 ($\epsilon: 0.40 \rightarrow 0.05$) & $+0.034 \pm 0.135$ & $+0.040$ & $0.345$ & $+0.302$ & $0.525 \pm 0.112$ & $0.517 \pm 0.110$ \\
\bottomrule
\end{tabular}

\begin{tablenotes}[flushleft]
\item \textit{Note.} $n=20$ conversations per condition. $\Delta Q$ = quality improvement from Exchange~1 to Exchange~15; $\bar{Q}$ = average quality across all conversation exchanges; $Q_{\text{final}}$ = quality at Exchange~15. 
Statistical tests (columns 3 to 5) compare $\Delta Q$ between each RL configuration and the prior baseline using independent-samples $t$-tests. 
$^*p < 0.05$. Conventional benchmarks: medium effect $d=0.50$–$0.79$, large effect $d \geq 0.80$.
\end{tablenotes}

\end{threeparttable}
\end{sidewaystable}

%% file: AURA_phase_results_table.tex
\begin{sidewaystable}[htbp]
\centering
\begin{threeparttable}
\caption{Phase-Specific Quality Improvement by Condition.}
\label{tab:AURA_phase_results}

\begin{tabular}{lccccc}
\toprule
\textbf{Phase} & \textbf{Prior} & \textbf{Config 1} & \textbf{Config 2} & \textbf{Config 3} & \textbf{Significant} \\
 & \textbf{Baseline} & ($\epsilon=0.15$) & ($\epsilon=0.30$) & ($\epsilon: 0.40{\rightarrow}0.05$) & \textbf{Differences} \\
\hline
Early (1--5) & $+0.058 \pm 0.169$ & $+0.069 \pm 0.174$ & $+0.053 \pm 0.101$ & $+0.066 \pm 0.142$ & n.s. \\
Mid (6--10) & $+0.021 \pm 0.128$ & $+0.004 \pm 0.193$ & $-0.016 \pm 0.117$ & $+0.041 \pm 0.157$ & n.s. \\
Late (11--15) & $+0.010 \pm 0.150$ & $-0.116 \pm 0.143$ & $+0.043 \pm 0.131$ & $-0.016 \pm 0.151$ & $p=0.025^*$ \\
\bottomrule
\end{tabular}

\begin{tablenotes}[flushleft]
\item \textit{Note.} Values represent mean quality change within each conversation phase, computed as the last-exchange quality minus the first-exchange quality within that phase, averaged across all conversations ($n=20$ per condition). Statistical tests compare each RL configuration with the prior baseline using independent-samples $t$-tests. Only Configuration~1 in the late phase achieved statistical significance. n.s. = not significant at $\alpha=0.05$.
\item Significance detail: $^*p<0.05$ for Config~1 vs. Prior in the late phase ($t(38)=-2.357$, $d=-0.862$), indicating significant quality degradation.
\end{tablenotes}

\end{threeparttable}
\end{sidewaystable}

%% file: AURA_action_dist_result_table.tex
\begin{sidewaystable}[htbp]
\centering
\begin{threeparttable}
\caption{Action Type Distribution across Conditions.}
\label{tab:AURA_action_dist}

\begin{tabular}{lccccc}
\toprule
\textbf{Action Type} & \textbf{Prior} & \textbf{Config 1} & \textbf{Config 2} & \textbf{Config 3} & \textbf{Change (C2)} \\
 & \textbf{Baseline} & ($\epsilon=0.15$) & ($\epsilon=0.30$) & ($\epsilon: 0.40{\rightarrow}0.05$) & \\
\midrule
Specification & 58.5\% & 25.4\% & 22.2\% & 23.6\% & $-36.3$pp \\
Elaboration   & 27.4\% & 53.3\% & 48.1\% & 51.4\% & $+20.7$pp \\
Topic Probe   & 12.9\% & 8.2\%  & 13.0\% & 14.9\% & $+0.1$pp  \\
Validation    & 0.8\%  & 2.9\%  & 7.9\%  & 3.8\%  & $+7.1$pp  \\
Continuation  & 0.4\%  & 10.2\% & 8.8\%  & 6.2\%  & $+8.4$pp  \\
\bottomrule
\end{tabular}

\begin{tablenotes}[flushleft]
\item \textit{Note.} pp = percentage point. Values represent percentage of total actions selected during 15-exchange conversations ($n=20$ conversations per condition). Prior baseline percentages closely match empirical frequencies from original TigerGPT data (Table~\ref{tab:AURA_action-distribution}: specification 62.3\%, elaboration 23.6\%, topic probe 12.8\%, validation 0.9\%, continuation 0.4\%), validating baseline implementation within expected sampling variability.
\end{tablenotes}
\end{threeparttable}
\end{sidewaystable}